\documentclass[useAMS,usenatbib]{mn2e}

\usepackage{graphicx}
\usepackage{amsmath}
\usepackage{amssymb}
\usepackage{color}
\usepackage{subfig}
\usepackage[breaklinks,colorlinks,citecolor=blue,linkcolor=red]{hyperref} 
\usepackage[all]{hypcap}

\newcommand\msun{\, \rm M_\odot}

\newcommand\kms{\, \rm km\,s^{-1}}

\newcommand\gyr{{\, \rm Gyr}}

\newcommand\eout{{e_{\rm out}}}
\newcommand\ein{{e_{\rm in}}}
\newcommand\aout{{a_{\rm out}}}
\newcommand\ain{{a_{\rm in}}}
\newcommand\amax{{a_{\rm 3, max}}}

\newcommand\mbha{{m_{\rm BH,1}}}
\newcommand\mbhb{{m_{\rm BH,2}}}
\newcommand\mbhc{{m_{\rm BH,3}}}
\newcommand\sigbh{{\sigma_{\rm BH}}}

\newcommand\vk{{v_\mathrm{k}}}
\newcommand\chieff{{\chi_{\rm eff}}}
\newcommand\chibh{{\chi_{\rm bh}}}

\title[Effective spin of BH mergers in triples]{Effective spin distribution of black hole mergers in triples}
\author[G. Fragione, B. Kocsis]{\parbox{\textwidth}{Giacomo Fragione$^{1,2}$\thanks{E-mail: giacomo.fragione@northwestern.edu}, Bence Kocsis$^{3}$}\\
$^1$Department of Physics \& Astronomy, Northwestern University, Evanston, IL 60202, USA\\
$^2$Center for Interdisciplinary Exploration \& Research in Astrophysics (CIERA), Evanston, IL 60202, USA\\
$^3$Institute of Physics, E\"{o}tv\"{o}s University, P\'azm\'{a}ny P. s. 1/A, Budapest, 1117, Hungary}

\hypersetup{draft}

\begin{document}

\maketitle

\begin{abstract}
Many astrophysical scenarios have been proposed to explain the several black hole (BH) and neutron star binary mergers observed via gravitational waves (GWs) by the LIGO-Virgo collaboration. Contributions from various channels can be statistically disentangled by mass, spin, eccentricity and redshift distributions of merging binaries. In this paper, we investigate the signatures of BH-BH binary mergers induced by a third companion through the Lidov-Kozai mechanism in triple systems. We adopt different prescriptions for the supernovae natal kicks and consider different progenitor metallicities and initial orbital parameters. We show that the typical eccentricity in the LIGO band is $0.01$-$0.1$ and that the merger rate is in the range $0.008-9 \ \mathrm{Gpc}^{-3}\ \mathrm{yr}^{-1}$, depending on the natal kick prescriptions and progenitor metallicity. Furthermore, we find that the typical distribution of effective projected spin is peaked at $\chi_{\rm eff}\sim 0$ with significant tails. We show that the triple scenario could reproduce the distribution of $\chi_{\rm eff}$. We find that the triple channel may be strongly constrained by the misalignment angle between the binary component spins in future detections with spin precession.
\end{abstract}


\section{Introduction}
\label{sect:intro}

Many astrophysical scenarios have been proposed to explain the several black hole (BH) and neutron star (NS) binary mergers observed via gravitational wave (GW) emission by the LIGO-Virgo collaboration. Possibilities include isolated binary evolution \citep{bel16b,kruc2018}, mergers catalyzed by stellar envelope expansion \citep{Tagawa2018}, mergers in star clusters \citep{askar17,baner18,frak18,rod18}, mergers in galactic nuclei \citep{olea09,antoper12,fragrish2018,grish18,hamil2019,rass2019}, mergers in gaseous disks \citep{Bartos2017,Stone2017,Tagawa2019}, and Lidov-Kozai (LK) mergers in isolated triple and quadruple systems \citep{ant17,sil17,arc2018,fragk2019,liu2019}.

With the improving sensitivity of LIGO-Virgo and the expected commissioning of KAGRA and LIGO India, hundreds of detections of merging systems is expected within the decade. Thus, it is fundamental to provide tools to distinguish among the mergers that originate in different astrophysical channels. It has been shown that useful parameters that can help doing so are the masses, spins, eccentricity and redshift of the merging binaries. Their distributions can be used as an indicator to statistically disentangle among the contributions of the several scenarios. This includes identifying correlations between mass and eccentricity
\citep{Breiviketal2016,gondan2018}, mass and spins \citep{PostnovKuranov2017,ArcaSedaBenacquista2018,Yangetal2019b}, 
the mass distribution \citep{Stevensonetal2015,olea16,Fishbachetal2017,Zevinetal2017,Mandeletal2017,Kocsisetal2018,perna2019,Yangetal2019a}, the spin distribution \citep{Fishbachetal2017}, eccentricity distribution in the LIGO/Virgo band \citep{wen03,olea09,Cholisetal2016,arc2018,gondan2018,Samsingetal2018,zevin18} and in the LISA band \citep{OLearyetal2006,Nishizawaetal2016,Nishizawaetal2017,DOrazioSamsing2018,SamsingDorazio2018}, spin orientations \citep{rodze2016,Stevensonetal2017,TalbotThrane2017,vitale2017,LiuLai2017,LiuLai2018,Farretal2018,gerosa2018,LiuLai2018,Lopezetal2018}, the projected effective spin parameter \citep{antonini2018,Ngetal2018,Schoderetal2018,Zaldarriagaetal2018}, 
and other waveform features \citep{Meironetal2017,Inayoshietal2017,Samsingetal2018,Kremeretal2018,sdk2019}.

\begin{table}
\caption{Total masses ($m_{\rm BH,1}+m_{\rm BH,2}$) and effective spin ($\chi_{\rm eff}$) of merging BH-BH binaries inferred by the LIGO-Virgo collaboration \citep{ligo2018} and the IAS group \citep{venu2019,zack2019}.}
\centering
\begin{tabular}{lcc}
\hline
Group & $m_{\rm BH,1}+m_{\rm BH,2}$ ($\msun$) & $\chi_{\rm eff}$\\
\hline\hline
LIGO-Virgo & 18.6$^{+6.9}_{-3.9}$ & 0.03$^{+0.19}_{-0.07}$\\
LIGO-Virgo & 21.4$^{+11.0}_{-5.7}$ & 0.18$^{+0.20}_{-0.12}$\\
LIGO-Virgo & 36.8$^{+19.0}_{-10.3}$ & 0.05$^{+0.31}_{-0.20}$\\
LIGO-Virgo & 50.8$^{+12.2}_{-10.2}$ & -0.04$^{+0.17}_{-0.21}$\\
LIGO-Virgo & 55.8$^{+8.4}_{-7.0}$ & 0.07$^{+0.12}_{-0.12}$\\
LIGO-Virgo & 58.8$^{+13.4}_{-11.1}$ & 0.08$^{+0.17}_{-0.17}$\\
LIGO-Virgo & 62.1$^{+11.8}_{-9.9}$ & -0.09$^{+0.18}_{-0.21}$\\
LIGO-Virgo & 66.2$^{+7.7}_{-7.5}$ & -0.01$^{+0.12}_{-0.13}$\\
LIGO-Virgo & 68.5$^{+17.9}_{-14.5}$ & 0.09$^{+0.22}_{-0.26}$\\
LIGO-Virgo & 84.2$^{+25.3}_{-20.3}$ & 0.37$^{+0.21}_{-0.25}$\\
\hline\hline
IAS & 23.2$^{+8.55}_{-8.55}$ & 0.25$^{+0.19}_{-0.19}$\\
IAS & 23.3$^{+6.02}_{-6.02}$ & 0.27$^{+0.11}_{-0.11}$\\
IAS & 39.3$^{+9.81}_{-9.81}$ & 0.05$^{+0.14}_{-0.14}$\\
IAS & 44.1$^{+9.69}_{-9.69}$ & -0.16$^{+0.21}_{-0.21}$\\
IAS & 49.7$^{+5.94}_{-5.94}$ & -0.09$^{+0.09}_{-0.09}$\\
IAS & 52.7$^{+9.34}_{-9.34}$ & 0.79$^{+0.11}_{-0.11}$\\
IAS & 55.6$^{+7.06}_{-7.06}$ & -0.30$^{+0.17}_{-0.17}$\\
IAS & 55.8$^{+4.73}_{-4.73}$ & 0.05$^{+0.07}_{-0.07}$\\
IAS & 59.6$^{+7.70}_{-7.70}$ & 0.08$^{+0.12}_{-0.12}$\\
IAS & 62.2$^{+6.38}_{-6.38}$ & -0.05$^{+0.12}_{-0.12}$\\
IAS & 65.1$^{+4.84}_{-4.84}$ & -0.05$^{+0.07}_{-0.07}$\\
IAS & 67.9$^{+8.07}_{-8.07}$ & 0.09$^{+0.13}_{-0.13}$\\
IAS & 69.1$^{+9.07}_{-9.07}$ & -0.09$^{+0.20}_{-0.20}$\\
IAS & 74.4$^{+10.5}_{-10.5}$ & 0.19$^{+0.19}_{-0.19}$\\
IAS & 76.5$^{+15.0}_{-15.0}$ & 0.05$^{+0.26}_{-0.26}$\\
IAS & 78.1$^{+11.2}_{-11.2}$ & -0.63$^{+0.23}_{-0.23}$\\
IAS & 84.6$^{+12.5}_{-12.5}$ & 0.43$^{+0.13}_{-0.13}$\\
\hline
\end{tabular}
\label{tab:data}
\end{table}

\begin{table*}
\caption{Models parameters: name, dispersion of BH kick-velocity distribution ($\sigbh$), progenitor metallicity ($Z$), maximum outer semi-major axis of the triple ($\amax$), fraction of stable triple systems after SNe ($f_{\rm stable}$), fraction of stable systems that merge from the $N$-body simulations ($f_{\rm merge}$).}
\centering
\begin{tabular}{lccccc}
\hline
Name & $\sigma$ ($\kms$) & $Z$ & $\amax$ (AU) & $f_{\rm stable}$ & $f_{\rm merge}$\\
\hline\hline
A1 & $260$ & $0.01$   & $2000$ & $7.8\times 10^{-5}$ & $0.06$\\
A2 & $100$ & $0.01$   & $2000$ & $1.5\times 10^{-3}$ & $0.06$\\
A3 & $0$   & $0.01$   & $2000$ & $3.1\times 10^{-2}$ & $0.08$\\
B1 & $260$ & $0.0001$ & $2000$ & $1.3\times 10^{-1}$ & $0.06$\\
B2 & $260$ & $0.001$  & $2000$ & $6.2\times 10^{-2}$ & $0.06$\\
B3 & $260$ & $0.005$  & $2000$ & $1.4\times 10^{-3}$ & $0.05$\\
B4 & $260$ & $0.015$  & $2000$ & $1.5\times 10^{-5}$ & $0.06$\\
C1 & $260$ & $0.01$   & $5000$ & $5.8\times 10^{-5}$ & $0.06$\\
C2 & $260$ & $0.01$   & $7000$ & $4.2\times 10^{-5}$ & $0.05$\\
\hline
\end{tabular}
\label{tab:models}
\end{table*}

GW emission is highly efficient at circularizing the orbit of an inspiraling binary. As a consequence, BHs that merge in isolation are expected to enter the LIGO frequency band ($10$ Hz) almost circular, with typical eccentricities in the range $e_{10\rm Hz}\sim 10^{-7}$--$10^{-6}$. In the case the BH binary merges in a hierarchical system as a result of the LK mechanism, a number of authors showed that the typical eccentricity has much higher values, in the range $\sim 10^{-2}$--$10^{-1}$ \citep{wen03,antchrod2016,frbr2019,fragk2019,frl2019}. If the BH binary is dynamically assembled in a cluster environment, the spectrum of possible eccentricities is rich, with three possible outcomes \citep{SamsingDorazio2018,zevin18}: (i) binaries that are ejected and merge outside the cluster have eccentricities $\sim 10^{-7}$--$10^{-6}$, as in the isolated binary case; (ii) binaries that merge as a result of a GW capture process have eccentricities $\sim 10^{-2}$--$10^{-1}$, as in the LK-induced mergers; (iii) binaries that merge within the cluster have intermediate eccentricities $\sim 10^{-5}$--$10^{-3}$. The recent non-detection of eccentric sources by LIGO-Virgo was used to place an upper limit on the rate of eccentric mergers with $e_{10\rm Hz}>0.1$ of $\mathcal{R}\leq 300\,f_{\rm ecc}\,\rm Gpc^{-3}\rm yr^{-1}$, where $f_{\rm ecc}$ is the fraction of mergers with $e_{10\rm Hz}>0.1$ \citep{lvs2019}. In comparison, the rate density of circular mergers is between $28-104\,\rm Gpc^{-3}\rm yr^{-1}$ for a power-law mass distribution prior \citep{ligo2018}. The eccentricity of dynamically formed BH binaries is much higher in the GW frequency band of LISA. 

Another powerful quantity to discriminate among the contributions of different astrophysical merger channels is $\chi_{\rm eff}$, the effective spin of the BH binary at merger defined as the mass-weighted average of the binary components' spins projected onto the orbital angular momentum vector of the binary. Binaries that evolve and merge in isolation are expected to have spin vectors that are aligned with the orbital angular momentum vector of the binary mainly due to tidal dissipation effects \citep{vitale2017,gerosa2018}. BHs that merge in star clusters are instead expected to have spin vectors distributed isotropically, as a consequence of the fact that the dynamical binary assembly does not prefer any specific direction \citep{rodze2016,arcaben2019}. For what concerns the focus of this work, LK-induced mergers in triple systems, \citet{antonini2018} and \citet{rodant2018} claimed that the BH binary component spins typically align in the perpendicular direction with respect to the orbital angular momentum of the binary due to the tertiary companion in a triple system, implying near-zero effective spins at merger. However, even more recently, \citet{liulai2019} have shown that these findings are sensitive to the applied approximations and orbit-averaging in the equations of motion. They showed that the BH spin exhibits a wide range of evolutionary paths, and different distributions of final spin-orbit misalignments can be produced depending on the system parameters.

In this paper, we study the dynamical evolution of BH triples by means of high-precision direct $N$-body simulations, including post-Newtonian (PN) terms up to 2.5PN order. We start from the main sequence progenitors of the BHs and model the supernova (SN) events that lead to the formation of the BH triple. We adopt different prescriptions for the SN natal kicks, and consider different progenitor metallicities and orbital parameters. We determine the expected distributions of various properties of merging systems, including masses, eccentricities, spin-orbit misalignment, and merger times.

The paper is organized as follows. In Section~\ref{sect:init}, we discuss the initial conditions adopted in this paper. In Section~\ref{sect:results}, we discuss the distribution of masses, orbital parameters, eccentricities and spin-orbit misalignements of merging systems. Finally, in Section~\ref{sect:conc}, we discuss the implications of our findings and draw our conclusions.

\section{Initial conditions}
\label{sect:init}

The stellar triples in our simulations are initialized as described in what follows. In total, we consider nine different models (see Table~\ref{tab:models}).

We consider a triple system comprised of an inner binary of mass $m_{\rm in}=m_1+m_2$ and a third body of mass $m_3$ that orbits the inner binary. The semi-major axis and eccentricity of the inner orbit are $\ain$ and $\ein$, respectively, and the semi-major axis and eccentricity of the outer orbit are $\aout$ and $\eout$, respectively. The inner and outer orbital plane have initial relative inclination $i_0$. 

We sample the mass $m_1$ of the most massive star in the inner binary from an initial mass function
\begin{equation}
\frac{dN}{dm} \propto m^{-\beta}\ ,
\label{eqn:bhmassfunc}
\end{equation}
in the mass range $20\msun$-$150\msun$, reflecting the progenitor of the BH. We fix $\beta=2.3$ \citep{kroupa2001}. We adopt a flat mass ratio distribution for the inner binary, $m_2/m_1$, and the outer binary, $m_3/(m_1+m_2)$ \citep{sana12,duch2013,sana2017}. The mass of the secondary in the inner binary and of the third companion are sampled from the range $20\msun$-$150\msun$. For comparison, we also estimate how the final rate changes if the mass ratio distribution is assumed to be log-uniform \citep{sana13}\footnote{From observations, \citet{duch2013} found that $f(q)\propto q^{1.16\pm0.16}$ and $q^{-0.01\pm0.03}$ for solar type stars with period less than or larger than $10^{5.5}\,$day, respectively, while \citet{sana13} found $f(q)\propto q^{-1.0\pm 0.4}$ for massive O-type stars.}.

We take the distribution of the inner and outer semi-major axis, $\ain$ and $\aout$, respectively, log-uniform \citet{kob2014}. In order to make sure that the periapsis distance of the inner binary is large enough that no common-envelope phase or mass transfer occurs\footnote{This would also shrink the inner orbit and make the LK mechanism strongly suppressed by the relativistic precession \citep[see e.g.][]{nao16}.}, we set a minimum orbital separation $\ain(1-\ein^2)\gtrsim 10$ AU \citep{ant17}. We adopt three values for the maximum separation of the triple respectively in different simulation sets, $\amax=2000$ AU, $3000$ AU, and $5000$ AU, see Table~\ref{tab:models} \citep{sana2014}. For the orbital eccentricities of the inner binary, $\ein$, and outer binary, $\eout$, we assume a flat distribution. The initial mutual inclination $i_0$ between the inner and outer orbit is drawn from an isotropic distribution, while the other relevant angles (i.e. argument of periapsis, argument of node, mean anomaly) are drawn randomly.

After sampling the relevant parameters, we check that the initial configuration satisfies the stability criterion of hierarchical triples of \citet{mar01}.

Given this set of initial conditions for the stellar triples, we assume that each of the three stars in the triple undergoes a SN event sequentially. We assume that SNe take place instantaneously (on a time-scale shorter than the orbital period), during which a given star looses mass instantaneously and collapses to a BH \citep[for details see e.g.][]{pijloo2012,toonen2016,frl2019}\footnote{We ignore the SN-shell impact on the companion stars.}. We determine the final masses of the BHs by using the fitting formulae to the results of the \textsc{parsec} stellar evolution code \citep[see Appendix C in][]{spera2015}. We adopt five different values of the progenitor metallicity as shown in Table~\ref{tab:models}, $Z=0.0001$, $0.001$, $0.005$, $0.01$, and $0.015$, which ultimately sets the final mass of the BH remnant.

As a result of the mass loss, the exploding star is imparted a kick to its center of mass \citep{bla1961}, and the system receives a natal kick due to recoil from an asymmetric supernova explosion. We assume that the BH natal velocity kick is drawn from a Maxwellian distribution
\begin{equation}
p(\vk)\propto \vk^2 e^{-\vk^2/\sigma^2}\ ,
\label{eqn:vkick}
\end{equation}
with a velocity dispersion $\sigma$. In our fiducial model, we consider $\sigma=260\kms$ for NSs, consistent with the distribution deduced by \citet{hobbs2005}. We run an additional model where we set $\sigma=100\kms$, consistent with the distribution of natal kicks found by \citet{arz2002}. We also adopt a model where no natal kick is imparted during BH formation (see Table~\ref{tab:models}). For BHs, we implement momentum-conserving kicks, in which we assume that the momentum imparted to a BH is the same as the momentum given to a NS \citep{fryer2001}. As a consequence, the kick velocities for the BHs is lowered by a factor of $1.4\msun/ m{_{\rm BH}}$ with respect to neutron stars. The value of $\sigma$ is highly uncertain. 

\begin{figure} 
\centering
\includegraphics[scale=0.55]{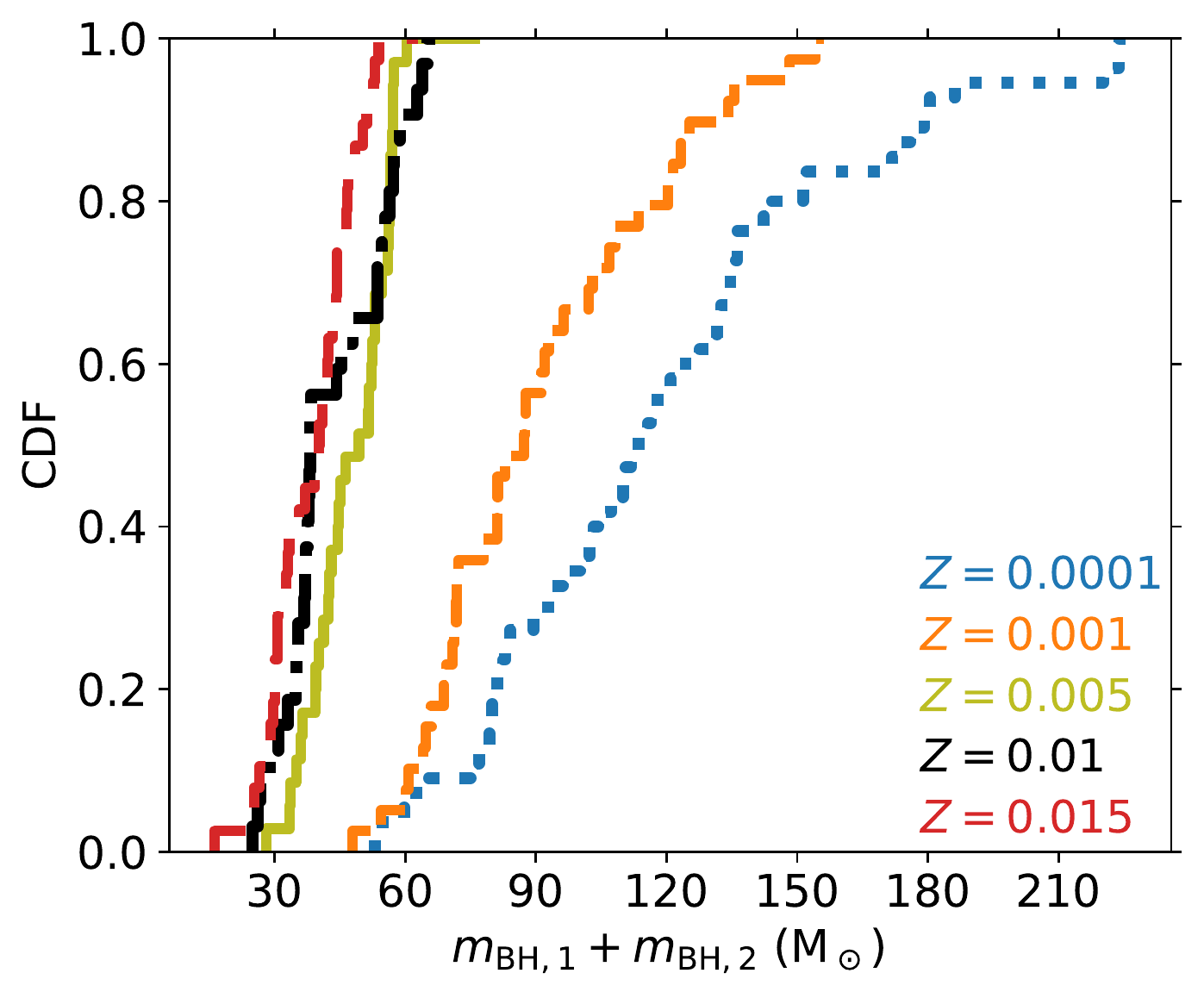}
\includegraphics[scale=0.55]{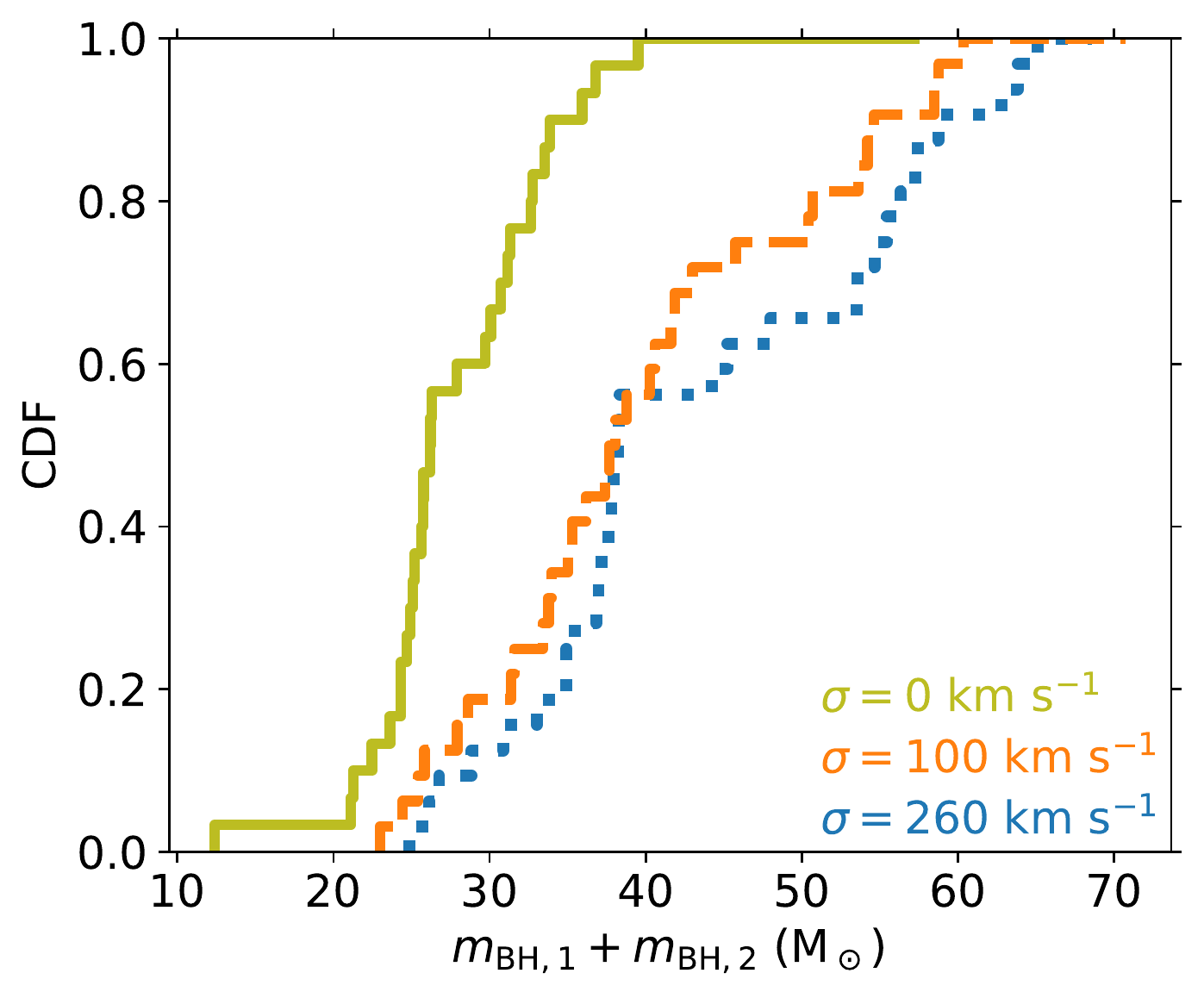}
\caption{Cumulative distribution function of the total BH mass of BH-BH binaries in triples that lead to a merger, for different values of progenitor metallicity $Z$ (top; $\sigma=260\kms$) and $\sigma$ (bottom; $Z=0.01$).} \label{fig:mass}
\end{figure}

After each SN event, the orbital elements of the triple are updated as appropriate \citep[see e.g.][]{flpk2019}, to account both for mass loss and natal kicks. The final masses of the BHs in the inner binary are $\mbha$ and $\mbhb$, and the outer companion $\mbhc$. We also check that the stability criterion of hierarchical triples of \citet{mar01} is satisfied and the triple is stable following mass loss and natal kicks. After all the SNe took place, we integrate the triple systems by means of the \textsc{ARCHAIN} code \citep{mik06,mik08}, including PN corrections up to order 2.5PN. We perform $\sim 750$--$1000$ simulations for each model in Table~\ref{tab:models}. We fix the maximum integration time as \citep{sil17},
\begin{equation}
T=\min \left(10^3 \times T_{\rm LK}, 10\ \gyr \right)\ ,
\label{eqn:tint}
\end{equation}
where $T_{\rm LK}$ is the triple LK timescale \citep{Antognini2015}
\begin{equation}
T_{\rm LK}=\frac{8}{15\pi}\frac{m_{\rm tot}}{\mbhc}\frac{P_{\rm out}^2}{P_{\rm in}}\left(1-e_{\rm out}^2\right)^{3/2}\ ,
\end{equation}
where $m_{\rm tot}=\mbha+\mbhb+\mbhc$ and $P_{\rm in}$ and $P_{\rm out}$ are the inner and outer orbital period, respectively. 

\section{Results}
\label{sect:results}

In our simulations, we find that $\sim 5\%$--$8\%$ of the triple systems lead to a merger of the inner binary. In the following subsections, we investigate the distribution of these mergers with respect to masses, orbital parameters, and spins.

\subsection{Mass distribution}

Figure~\ref{fig:mass} shows the cumulative distribution function (CDF) of the total BH mass of BH-BH binaries in triples that lead to a merger, for different values of $Z$ (top panel; $\sigma=260\kms$) and $\sigma$ (bottom panel; $Z=0.01$). Both $Z$ and $\sigma$ affect the distribution of total BH masses.

For $Z\lesssim 0.005$, the distributions do not depend significantly on the progenitor metallicity and the total BH mass is limited to $\sim 60\msun$. For higher metallicities, the progenitors can collapse to more massive BHs, up to $\sim 120\msun$--$140\msun$ \citep{spera2015}. We find that $\sim 50\%$ of the mergers have total mass $\gtrsim 90\msun$ and $\gtrsim 120\msun$ for $Z=0.001$ and $Z=0.0001$, respectively. This is consistent with the findings of \citet{floeb2019}. For comparison, we report the distribution of total masses inferred from LIGO-Virgo mergers and the IAS group \citep{ligo2018,venu2019,zack2019} in Table~\ref{tab:data}. We note that population of very massive binaries we find in our simulations may be due to the fact that we are using fitting formulae to single stellar evolutionary tracks to determine the BH mass from the mass of its progenitor \citep{spera2015}. We are not modelling the mass loss during neither possible episodes of Roche-lobe overflows nor possible common evolution phases. Nevertheless, we set in our initial conditions a minimum orbital separation $\ain(1-\ein^2)\gtrsim 10$ AU to ensure that the periapsis distance of the inner binary is large enough that no common-envelope phase or mass transfer occurs \citep{ant17}. These processes are modeled in binary systems, but not entirely comprehended in triple systems  \citep{rosa2019,hamd2019}.

\begin{figure} 
\centering
\includegraphics[scale=0.55]{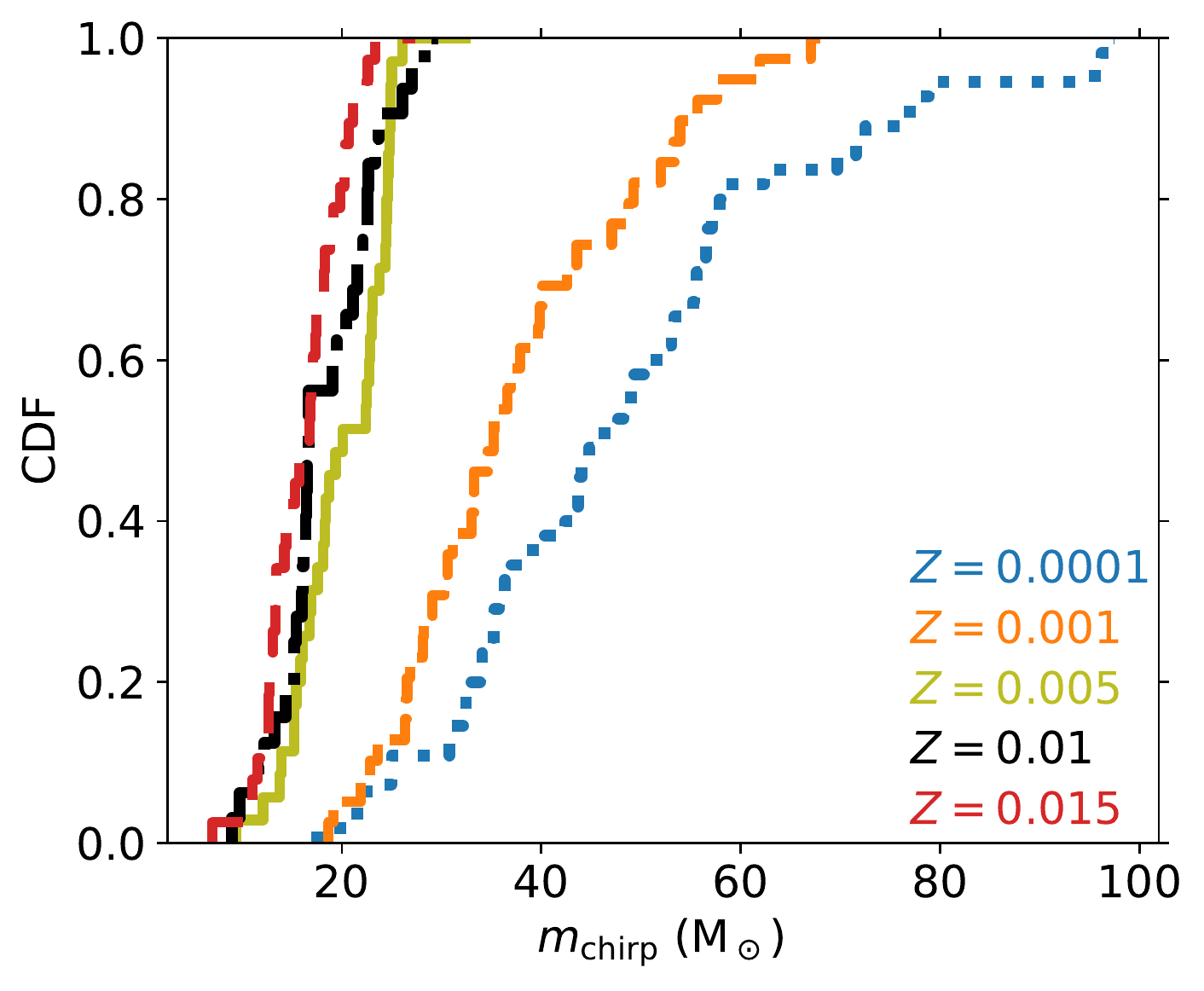}
\includegraphics[scale=0.55]{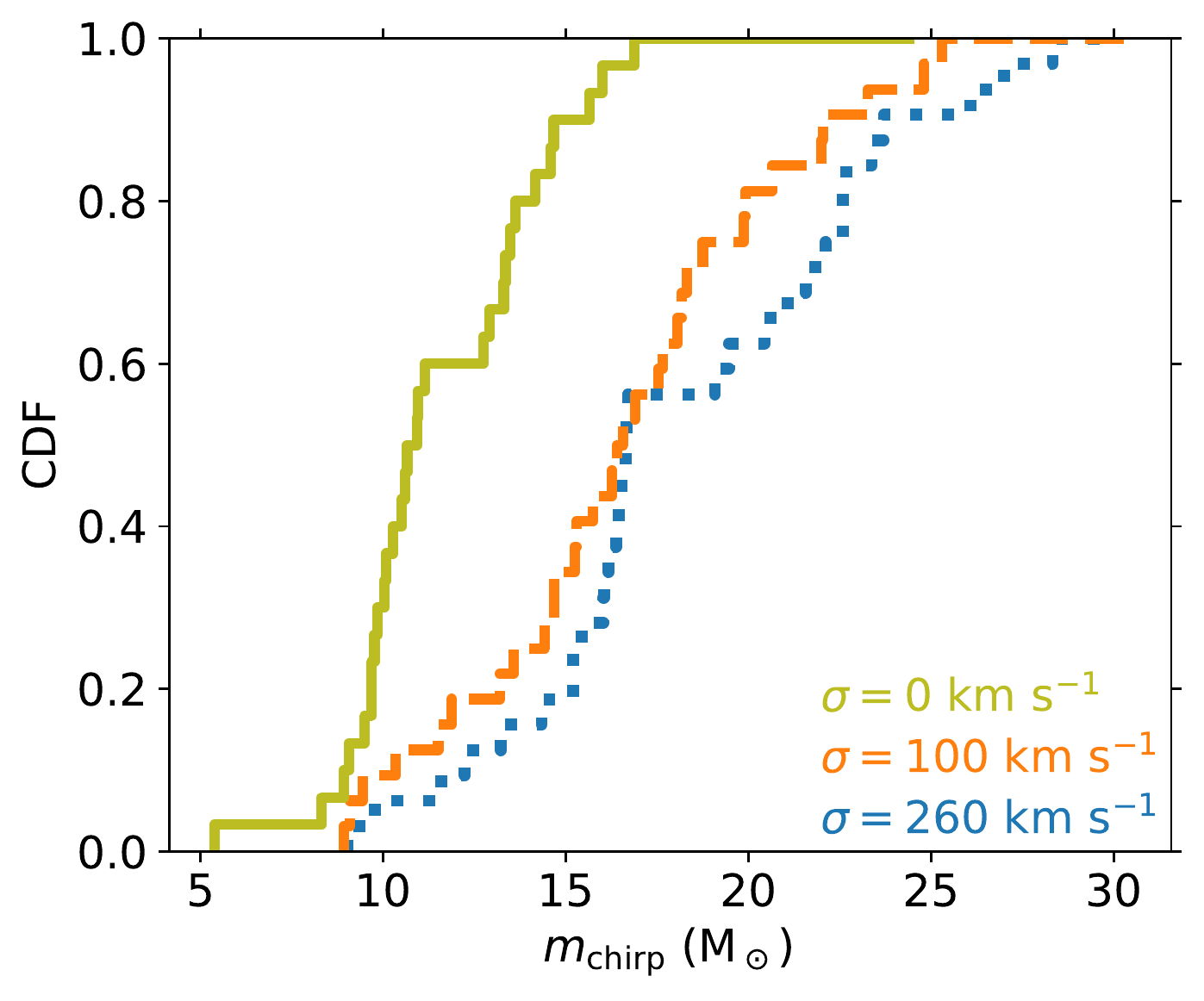}
\caption{Cumulative distribution function of the chirp mass of BH-BH binaries in triples that lead to a merger, for different values of $Z$ (top panel; $\sigma=260\kms$) and $\sigma$ (bottom panel; $Z=0.01$).}
\label{fig:mchirp}
\end{figure}

\begin{figure*} 
\centering
\includegraphics[scale=0.55]{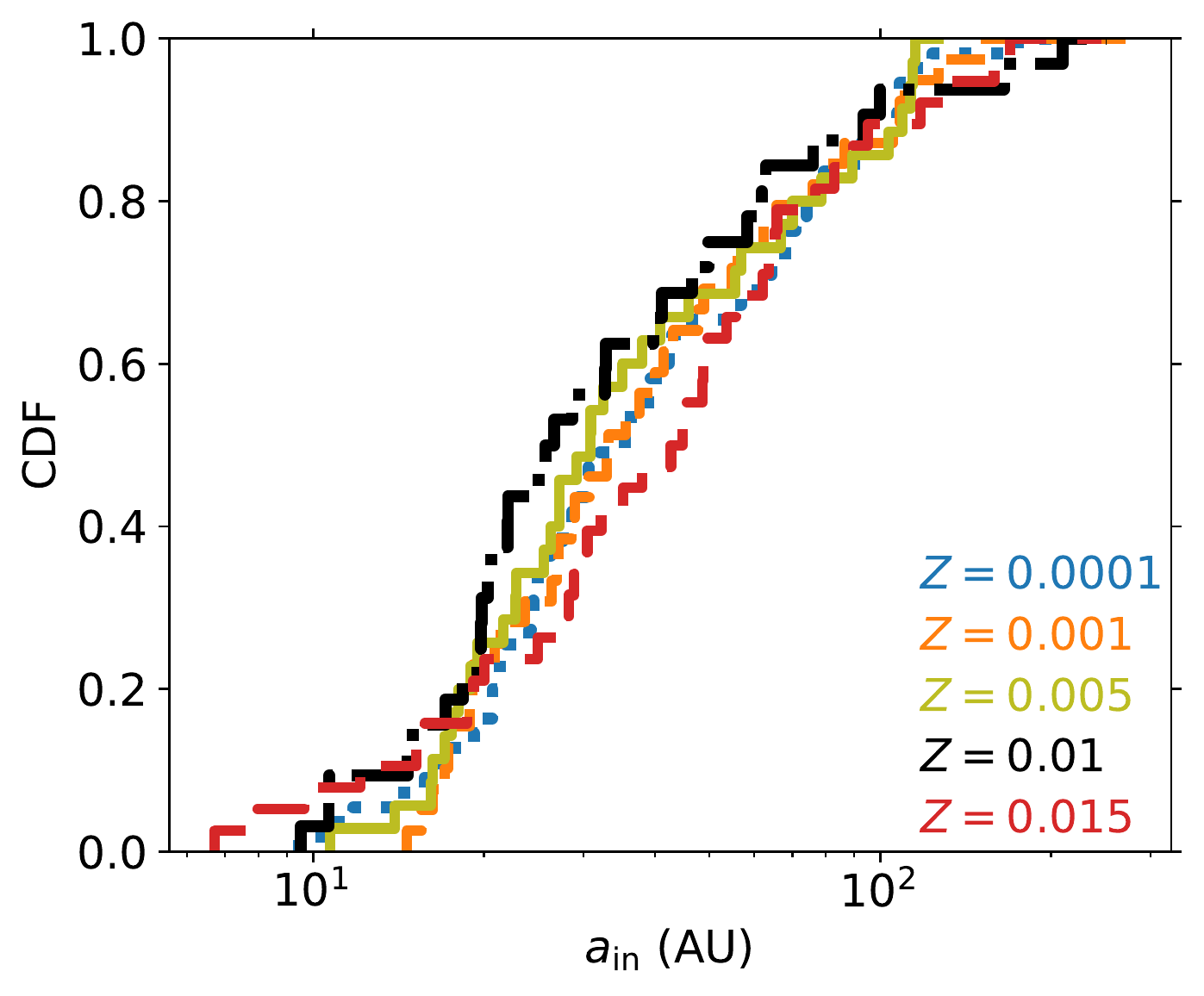}
\hspace{0.5cm}
\includegraphics[scale=0.55]{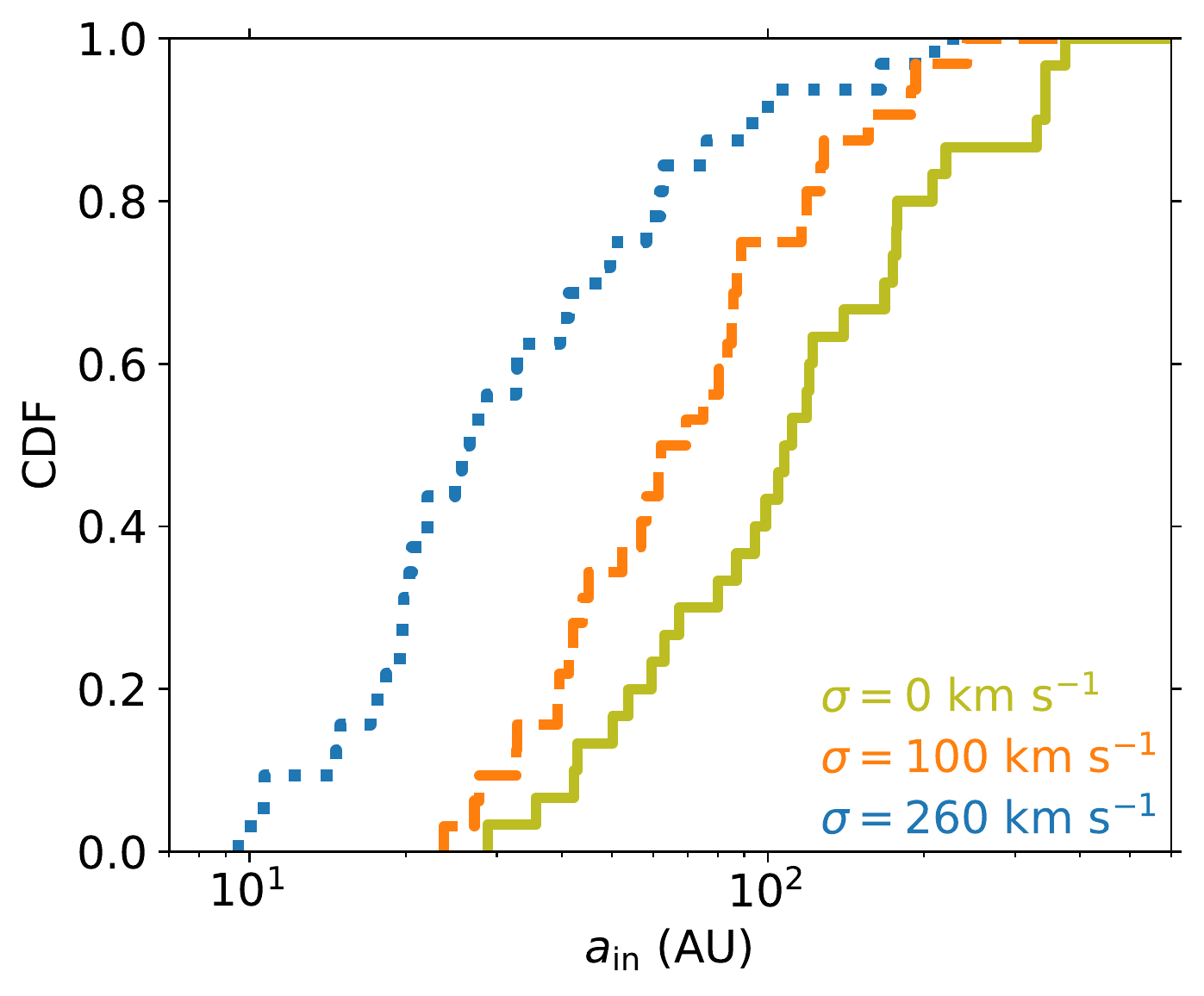}\\
\includegraphics[scale=0.55]{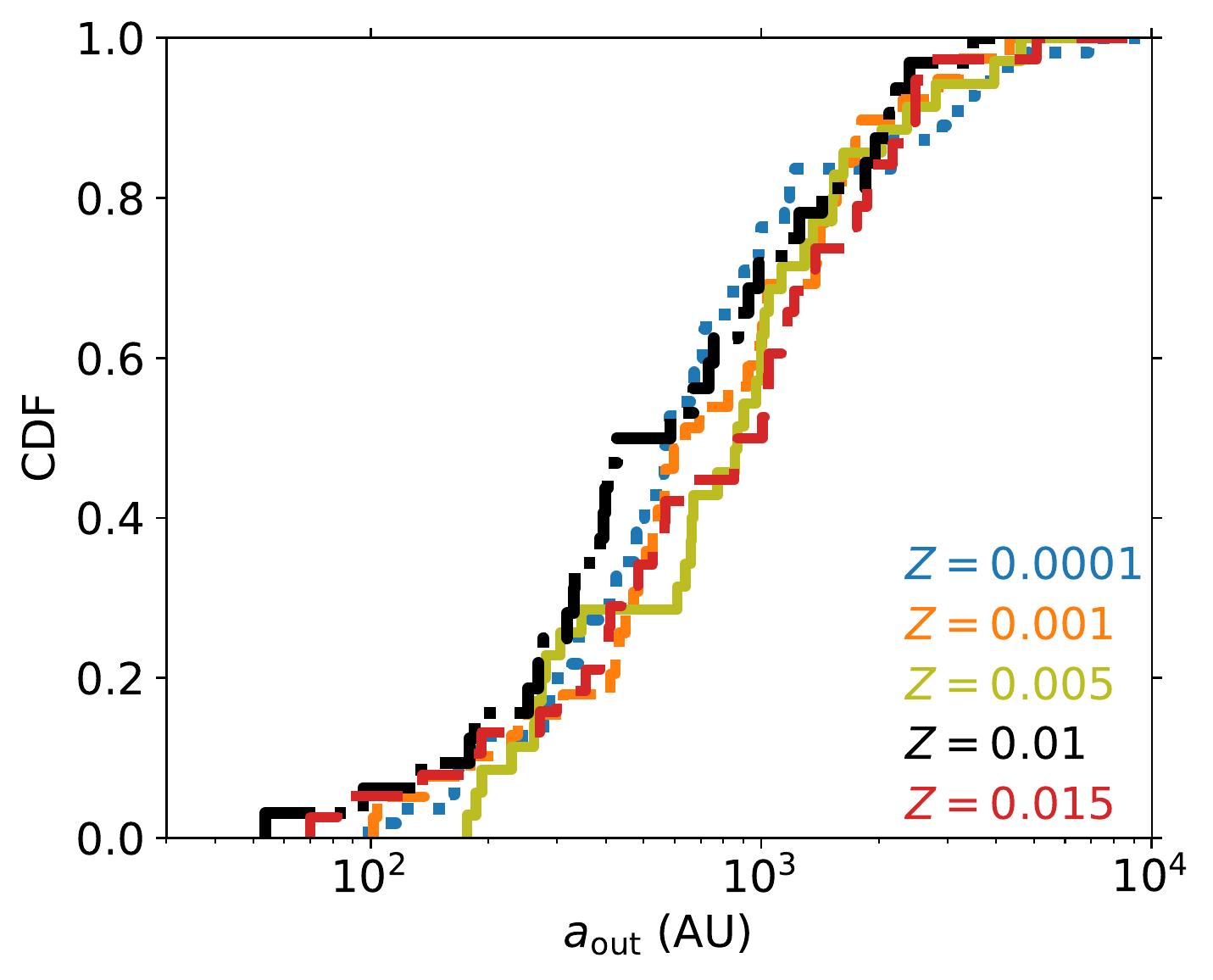}
\hspace{0.5cm}
\includegraphics[scale=0.55]{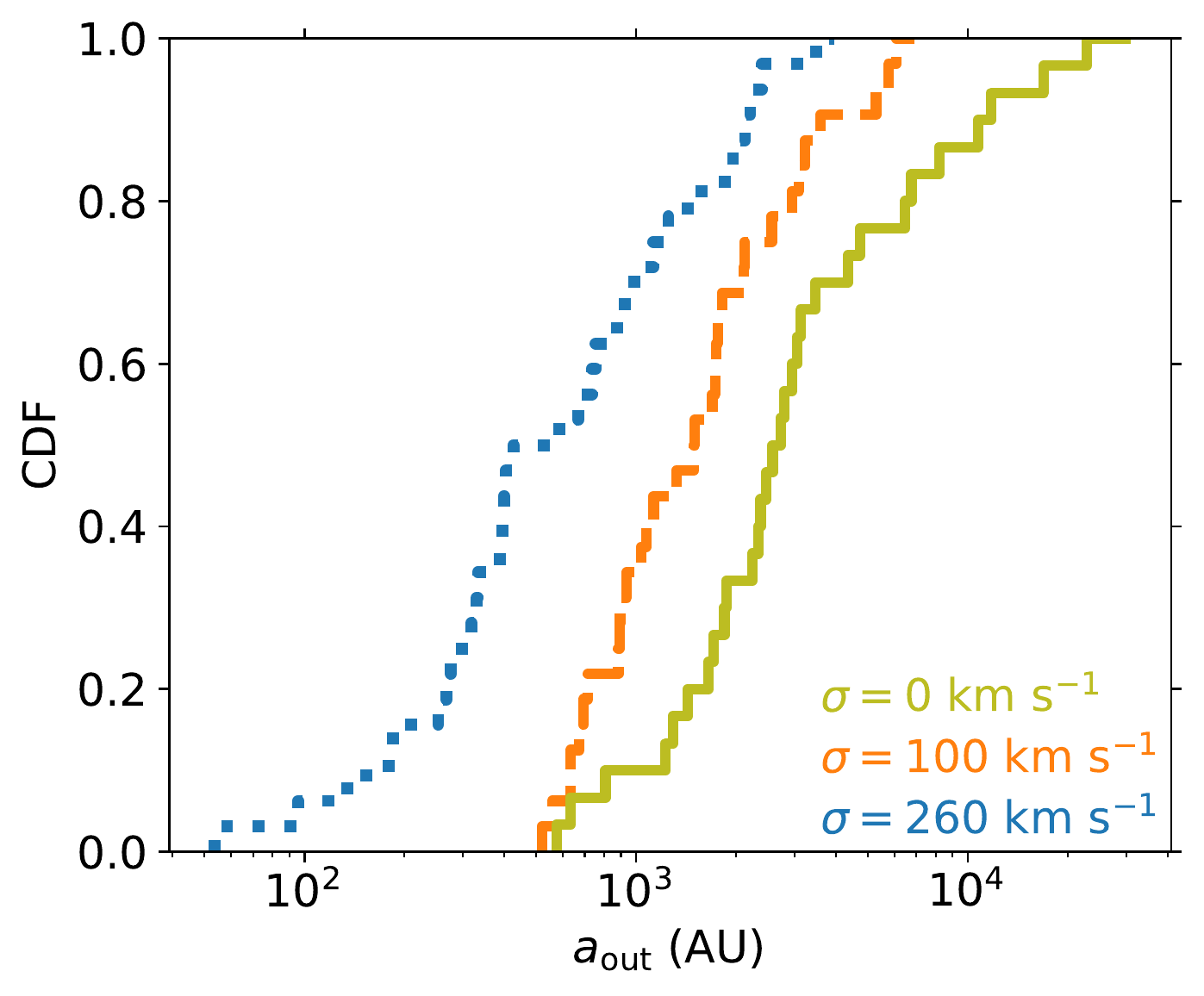}
\caption{Cumulative distribution function of inner (top) and outer (bottom) semi-major axis of BH-BH binaries in triples that lead to a merger. Left panel: $\ain$ and $\aout$ for different metallicities $Z$ and $\sigma=260\kms$; right panel: $\ain$ and $\aout$ for different values of $\sigma$ and $Z=0.01$.}
\label{fig:ainaout}
\end{figure*}

Concerning the role of $\sigma$, we find that the larger the kick velocity the larger the total BH mass on average. This is explained in relation to our assumption of momentum-conserving kicks, where $\sigbh\propto m_{\rm BH}^{-1}$. In our simulations, for $Z=0.01$, we find that $\sim 50\%$ of the mergers have total mass $\gtrsim 25\msun$ and $\gtrsim 40\msun$ for $\sigma=0\kms$ and $\sigma 100\kms$, respectively. Increasing the mean kick velocity to $\sigma=260\kms$ does not change considerably the distribution of total masses. Other prescriptions for the natal kicks may lead to different total mass distributions.

We report in Figure~\ref{fig:mchirp} the CDF of the chirp mass, the combination of masses measured most accurately during the inspiral,
\begin{equation}
m_\mathrm{chirp}=\frac{\left(\mbha\mbhb\right)^{3/5}}{\left(\mbha+\mbhb\right)^{1/5}}\ ,
\end{equation}
of BH-BH binaries in triples that lead to a merger, for different values of $Z$ (top panel; $\sigma=260\kms$) and $\sigma$ (bottom panel; $Z=0.01$). As in the case of the total mass, the distribution of chirp masses is not significantly affected by the progenitor metallicity for $Z\gtrsim 0.005$, $\sim 90\%$ of the BH-BH mergers have $m_{\rm chirp}\lesssim 20\msun$. We also find that $\sim 50\%$ fo the mergers have $m_{\rm chirp}\gtrsim 35\msun$ and $m_{\rm chirp}\gtrsim 45\msun$ for $Z=0.001$ and $Z=0.0001$, respectively. Moreover, lower natal kicks predict lower chirp masses. We find that $\sim 50\%$ of the BH-BH mergers have $m_{\rm chirp}\gtrsim 10\msun$ and $m_{\rm chirp}\gtrsim 18\msun$ for $\sigma=0\kms$ and $\sigma= 100\kms$, respectively. Furthermore, increasing the mean kick velocity to $\sigma=260\kms$ does not change the distribution of chirp masses significantly. This is consistent with the results of \citet{floeb2019}, who studied the BH-NS mergers in triples.

\subsection{Inner and outer semi-major axis}

Figure~\ref{fig:ainaout} illustrates the CDF of inner (top panel) and outer (bottom panel) semi-major axis of BH-BH binaries in triples that lead to a merger for different values of the progenitor metallicity ($\sigma=260\kms$) and mean natal kick velocities ($Z=0.01$). The progenitor metallicity does not affect the distribution of the inner and outer semi-major axes. On the other hand, the value of $\sigma$ highly affects their distribution, since higher kicks unbind wider triples. We find that $\sim 50$\% of the systems have $a_{\rm in}\lesssim 25$ AU, $\lesssim 60$ AU, $\lesssim 100$ AU for $\sigma=0\kms$, $100\kms$, $260\kms$, respectively, and $\sim 50$\% of the systems have $a_{\rm out}\lesssim 500$ AU, $\lesssim 1500$ AU, $\lesssim 2500$ AU for $\sigma=0\kms$, $100\kms$, $260\kms$, respectively. Similar conclusions were found in \citet*{flpk2019} and \citet{fmplk2019}.

\subsection{Eccentricity}

\begin{figure} 
\centering
\includegraphics[scale=0.55]{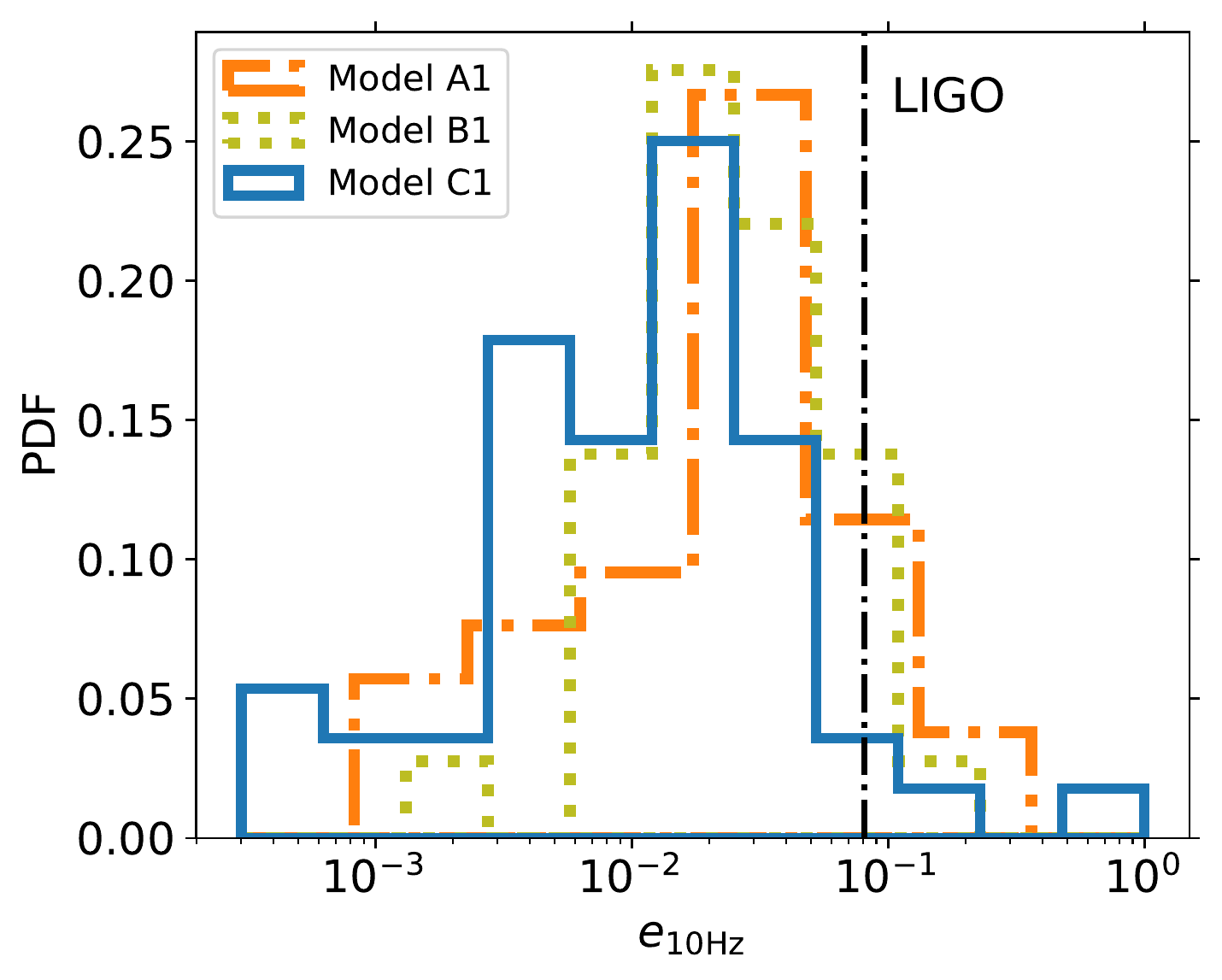}
\caption{Distribution of eccentricities at the moment the CO binaries enter the LIGO frequency band ($10$ Hz) for mergers produced by triples. The vertical line shows the conservative minimum $e_{\rm 10Hz}=0.081$ where LIGO/VIRGO/KAGRA network may distinguish eccentric sources from circular sources \citep{gond2019}. Note that for lower mass sources, the detection threshold at design sensitivity is $e_{\rm 10Hz}=0.023$, see footnote \ref{foot:ecc}).}
\label{fig:ecc}
\end{figure}

For the BH-BH binaries that merge in our simulations, we compute a proxy for the GW frequency, i.e. the frequency corresponding to the harmonic that gives the maximum GW emission \citep{wen03}
\begin{equation} 
f_{\rm GW}=\frac{\sqrt{G(\mbha+\mbhb)}}{\pi}\frac{(1+\ein)^{1.1954}}{[\ain(1-e_{\rm in}^2)]^{1.5}}\ .
\end{equation}
In Figure~\ref{fig:ecc}, we illustrate the probability distribution function (PDF) of eccentricities at the moment the BH binaries enter the LIGO frequency band, for Models A1, B1 and C1. We also show the minimum eccentricity $e_{\rm 10Hz}=0.081$ where the LIGO/VIRGO/KAGRA network may distinguish eccentric sources from circular sources \citep{gond2019}.\footnote{\label{foot:ecc}Note that the detection threshold of $e_{10\rm Hz}$ varies for different binary mass. For binaries with masses of those in the LIGO observing run O1 and O2, the minimum detectable eccentricity ranges between $e_{\rm 10Hz}=0.023$ and 0.081 \citep{gond2019}. } This value is consistent with the recent LIGO/VIRGO analysis \citep{lvs2019}.  A large fraction of merging BH-BH in triples retain a significant eccentricity at $10$ Hz, $\sim 42\%$ ($\sim 9\%$) of the merging systems have $e_{\rm 10Hz}$ higher than 0.023 (0.081). We note that a similar signature could be found for BH binaries that merge near supermassive and intermediate-mass black holes \citep{fragrish2018,frbr2019,flp2019}, in the GW capture scenario in star clusters \citep{sam2018}, and in triple \citep{ant17} and quadruple \citep{fragk2019} systems.

\subsection{Spin}

In our simulations, we incorporate the spin-orbit coupling effect by introducing
the spin vector ${\bf{S}}=S{\bf{\hat{S}}}$, where $S=(Gm_{\rm BH}^2/c)\chi$, where $0\le \chi\le 1$ is the dimensionless Kerr parameter. The leading order (1.5PN) de Sitter precession of the spins $\bf{S_1}$ and $\bf{S_2}$ of the BHs in the inner binary around the inner binary angular momentum $\bf{J}$ is given by \citep{Apostolatos1994}
\begin{eqnarray}
\frac{d\bf{S_{1}}}{dt}&=&{\bf{\Omega_1}}\times {\bf{S_{1}}}=\left[\frac{2G\mu}{c^2 r^3}\left(1+\frac{3m_2}{4m_1}\right){\bf{r}}\times {\bf{v}}\right] \times {\bf{S_{1}}} \\ 
\frac{d\bf{S_{2}}}{dt}&=&{\bf{\Omega_2}}\times {\bf{S_{2}}}=\left[\frac{2G\mu}{c^2 r^3}\left(1+\frac{3m_1}{4m_2}\right){\bf{r}}\times {\bf{v}}\right] \times {\bf{S_{2}}} \ ,
\end{eqnarray}
where $\mu=m_1m_2/(m_1+m_2)$ is the inner binary reduced mass, ${\bf{r}}={\bf{r_1}}-{\bf{r_2}}$ and ${\bf{v}}={\bf{v_1}}-{\bf{v_2}}$\footnote{We neglect the backreaction of $\bf{S_1}$ and $\bf{S_2}$ on $\bf{J}$ and the spin-spin precessional terms \citep{antonini2018,liulai2019}.}. GW measurements are especially sensitive to the following combination of the two spins \citep{LIGO2016,Vitale2017b}
\begin{equation}
\chieff=\frac{\mbha\chi_1\cos\theta_1+\mbhb\chi_2\cos\theta_2}{\mbha+\mbhb}\ ,
\end{equation}
where $\cos\theta_1=({\bf{\hat{S}_1}\cdot \bf{J}})/J$ and $\cos\theta_2=({\bf{\hat{S}_2}\cdot \bf{J}})/J$, where $\bf{J}$ is the total angular momentum of the inner BH binary, which is well approximated by the orbital angular momentum.

\begin{table}
\caption{Spin models: name, Kerr parameter of the BH ($\chi$), initial direction of the spins.}
\centering
\begin{tabular}{lcc}
\hline
Name & $\chi$ & Initial direction\\
\hline\hline
S1 & uniform              & $0^\circ \le\cos\theta_{1,2}^{\rm ini}\le 20^\circ$  \\
S2 & Eq.~\eqref{eqn:bhspin} & $0^\circ \le\cos\theta_{1,2}^{\rm ini}\le 20^\circ$ \\
T1 & uniform              & aligned with $\bf{J}$ \\
T2 & uniform              & $\cos\theta_{\rm 1,2}^{\rm ini}$ uniform \\
U1 & $0.2$                & $0^\circ \le\cos\theta_{1,2}^{\rm ini}\le 20^\circ$ \\
U2 & $0.5$                & $0^\circ \le\cos\theta_{1,2}^{\rm ini}\le 20^\circ$ \\
U3 & $0.8$                & $0^\circ \le\cos\theta_{1,2}^{\rm ini}\le 20^\circ$ \\
\hline
\end{tabular}
\label{tab:spins}
\end{table}

\begin{figure*} 
\centering
\includegraphics[scale=0.55]{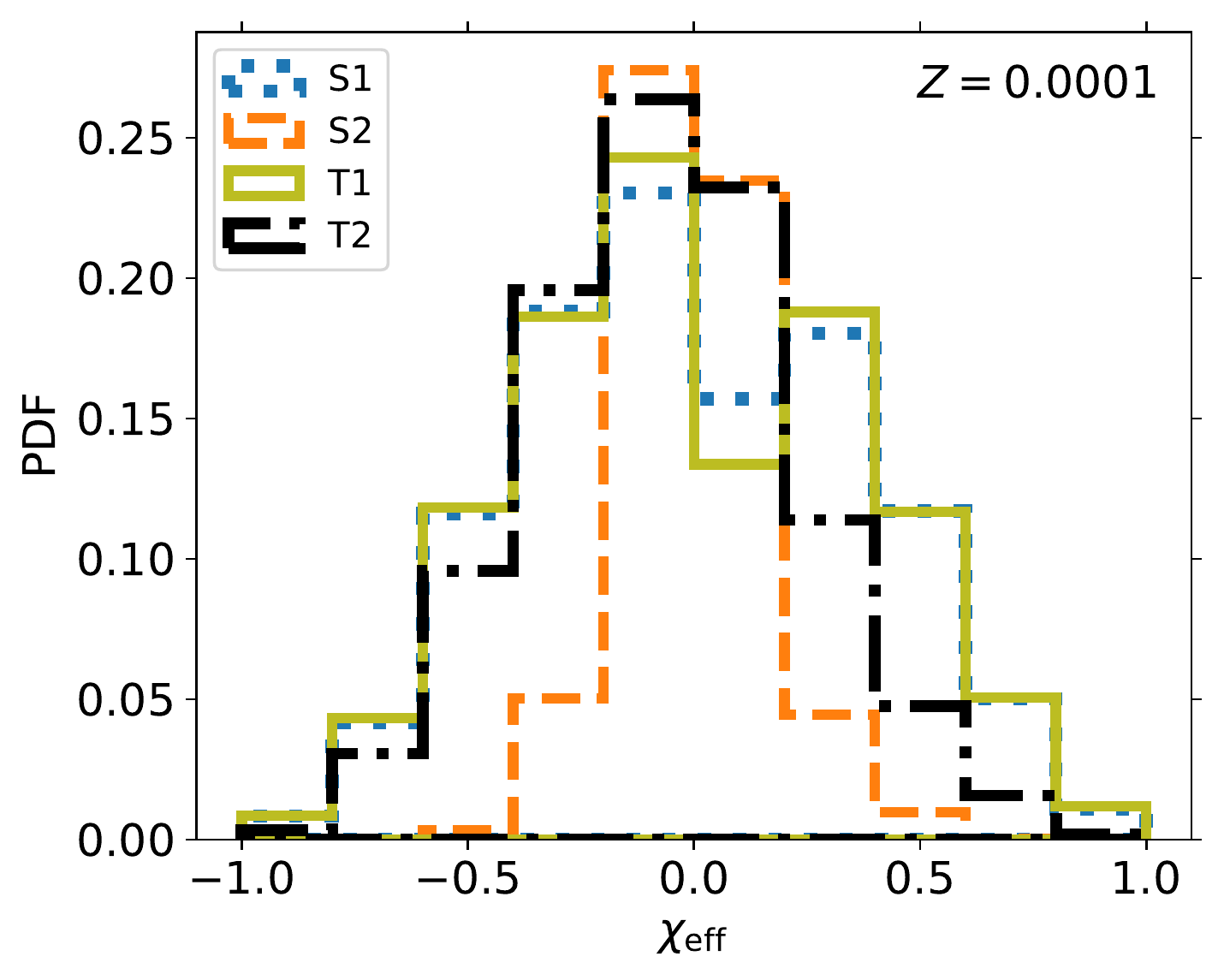}
\hspace{0.5cm}
\includegraphics[scale=0.55]{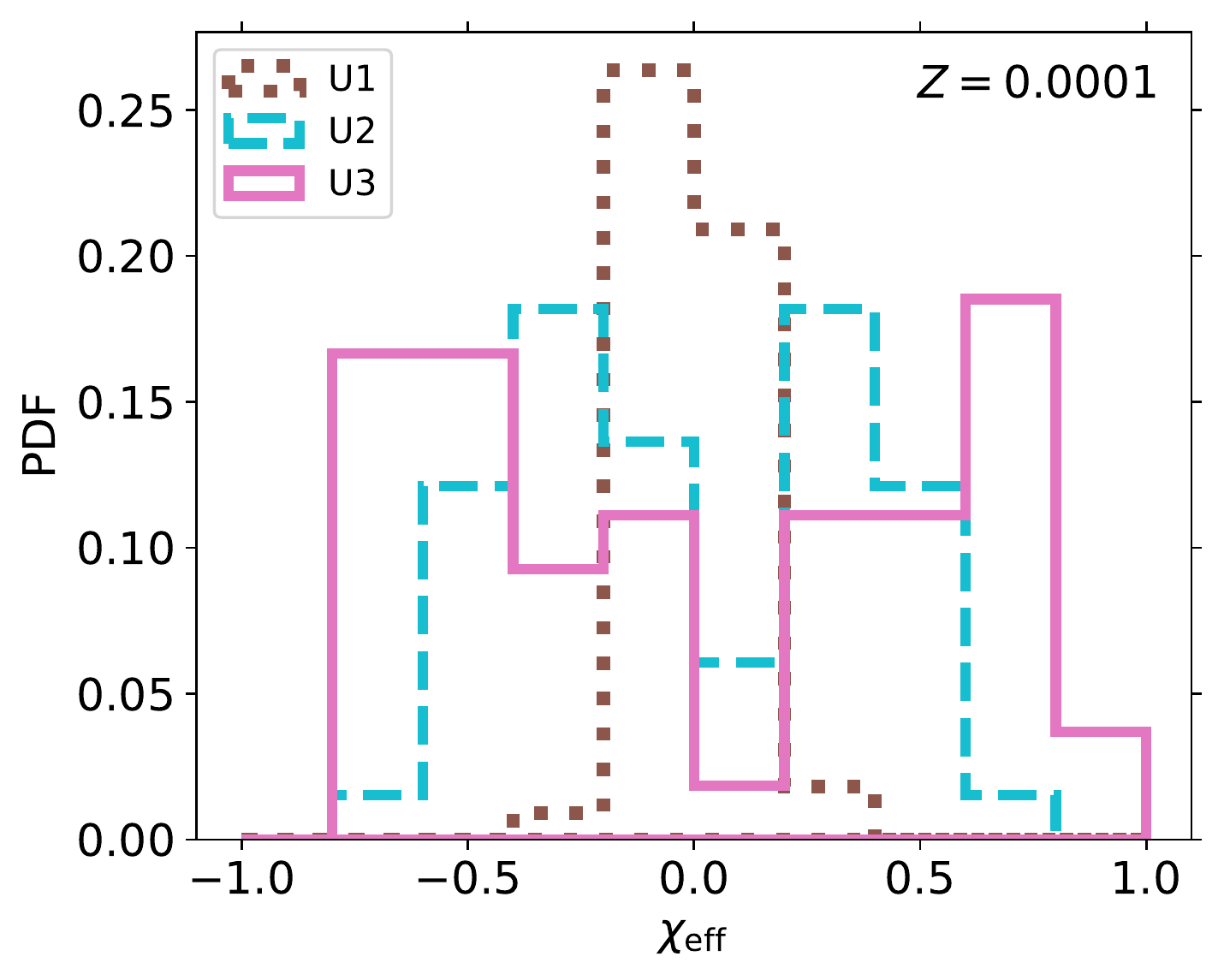}
\includegraphics[scale=0.55]{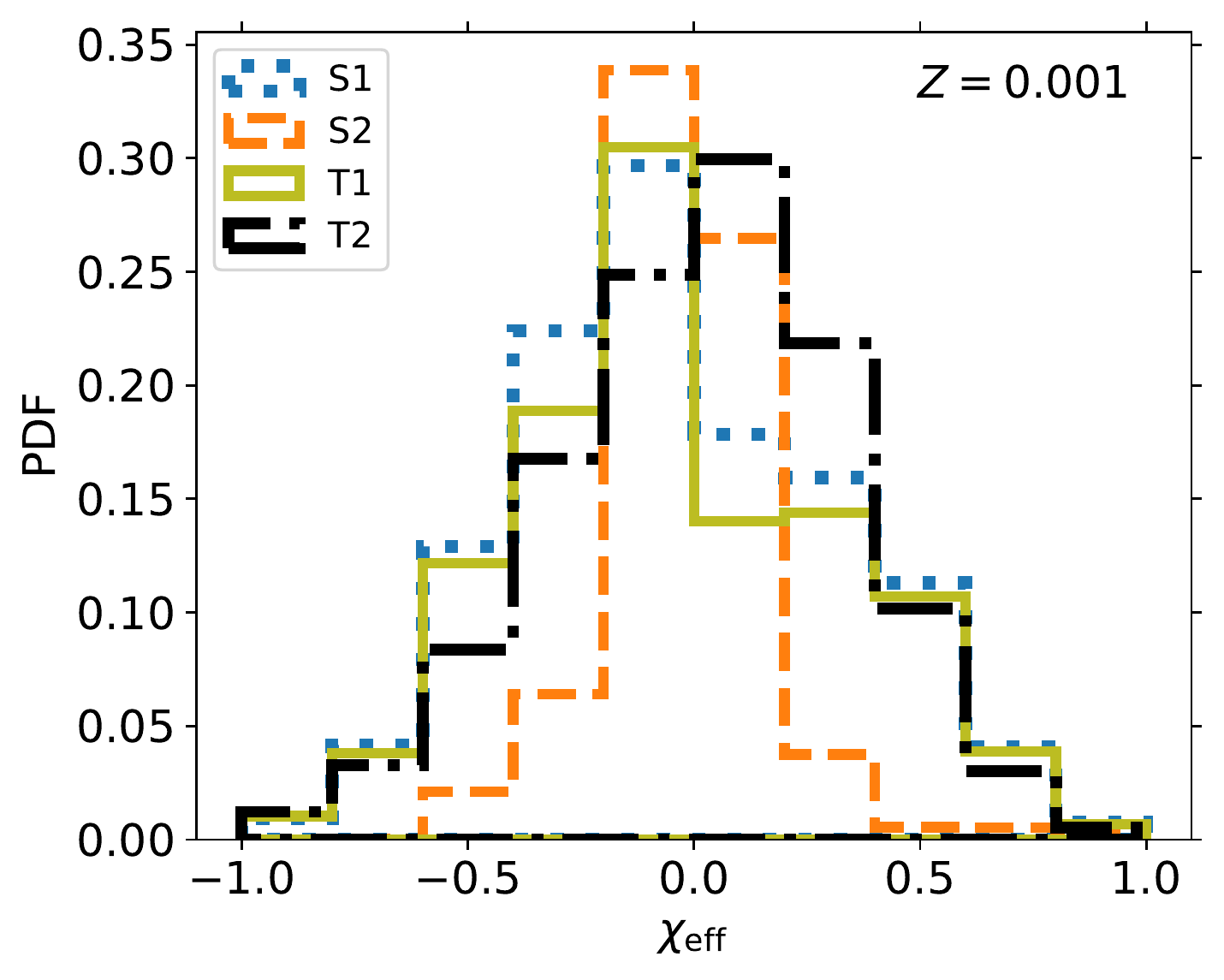}
\hspace{0.5cm}
\includegraphics[scale=0.55]{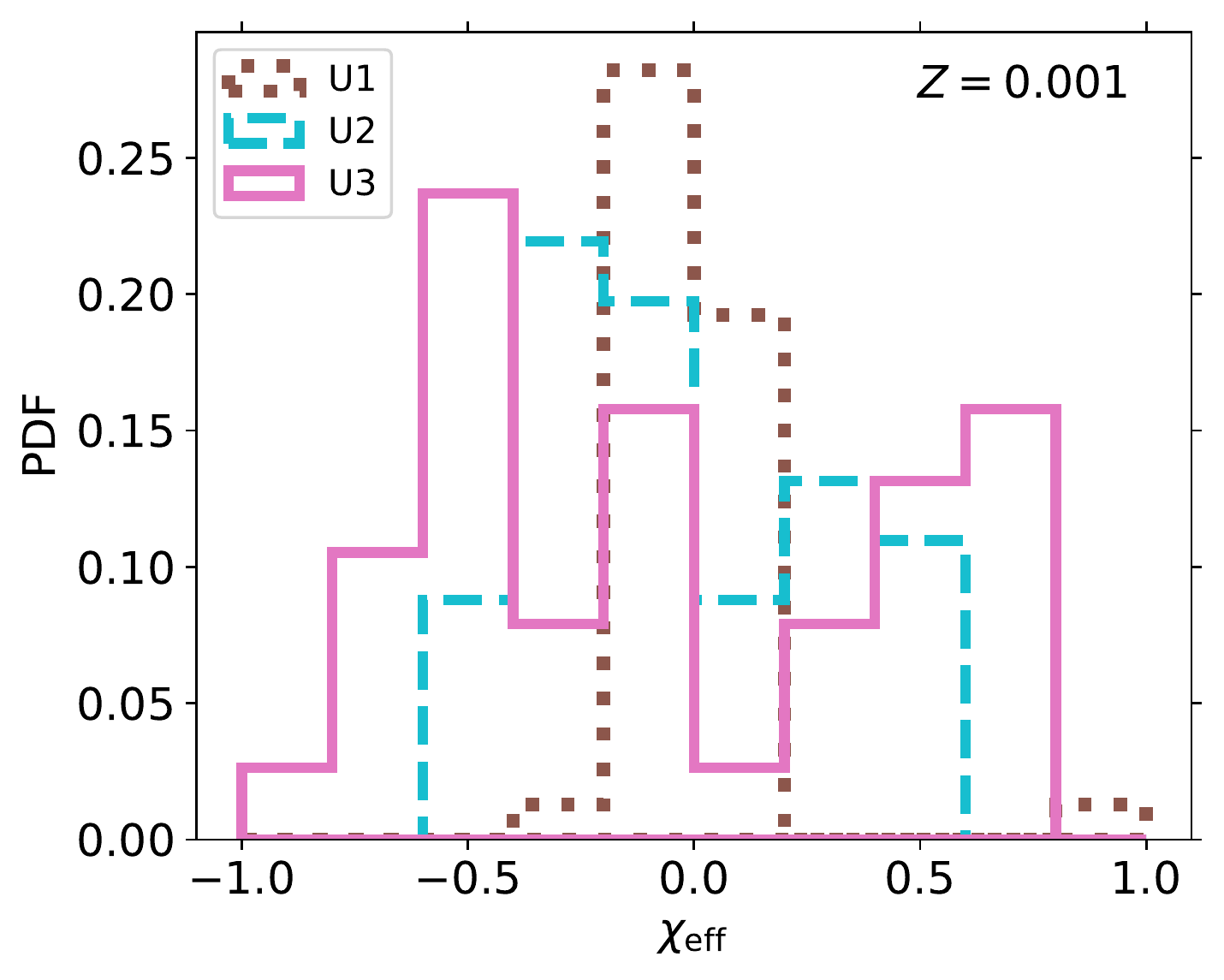}
\includegraphics[scale=0.55]{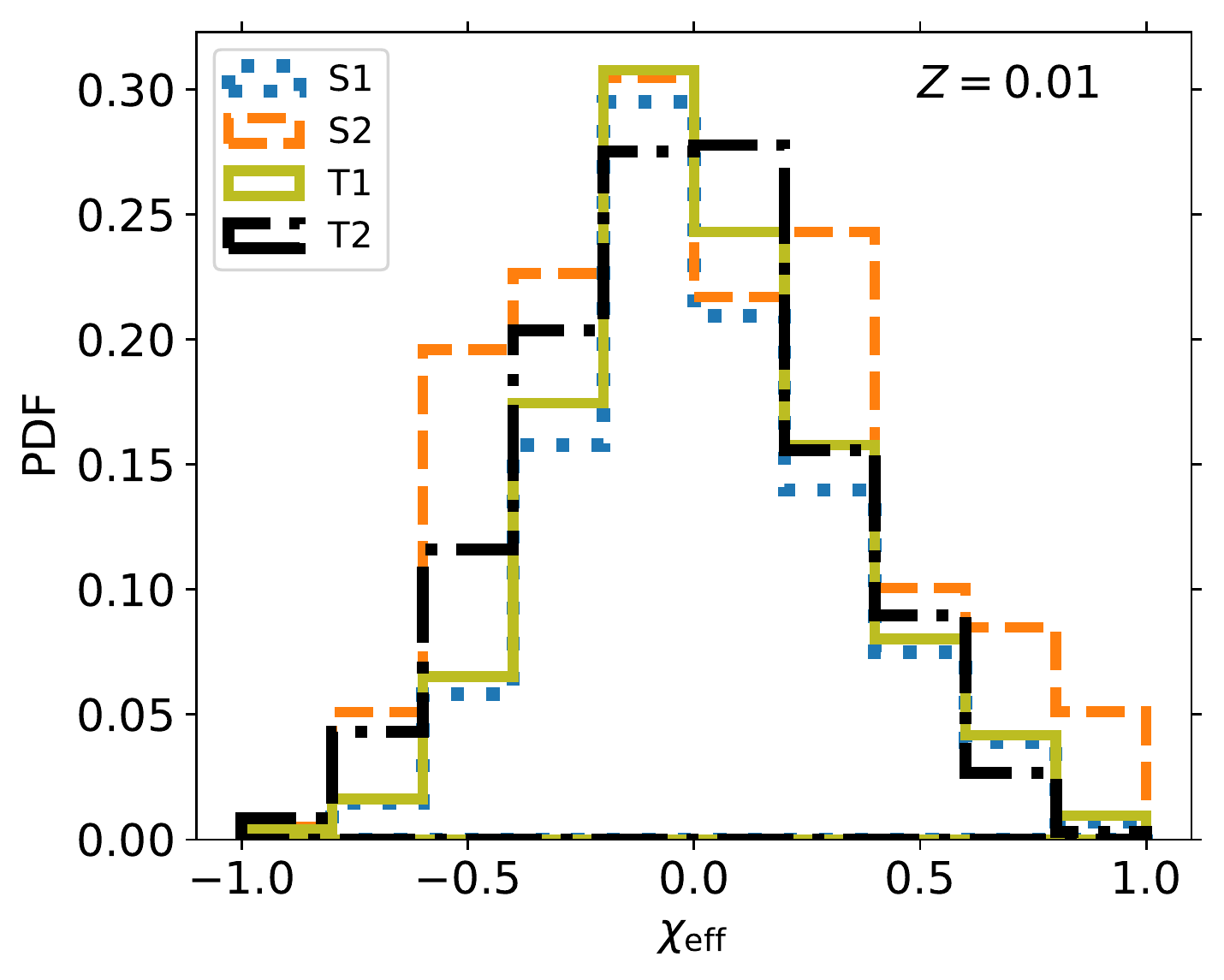}
\hspace{0.5cm}
\includegraphics[scale=0.55]{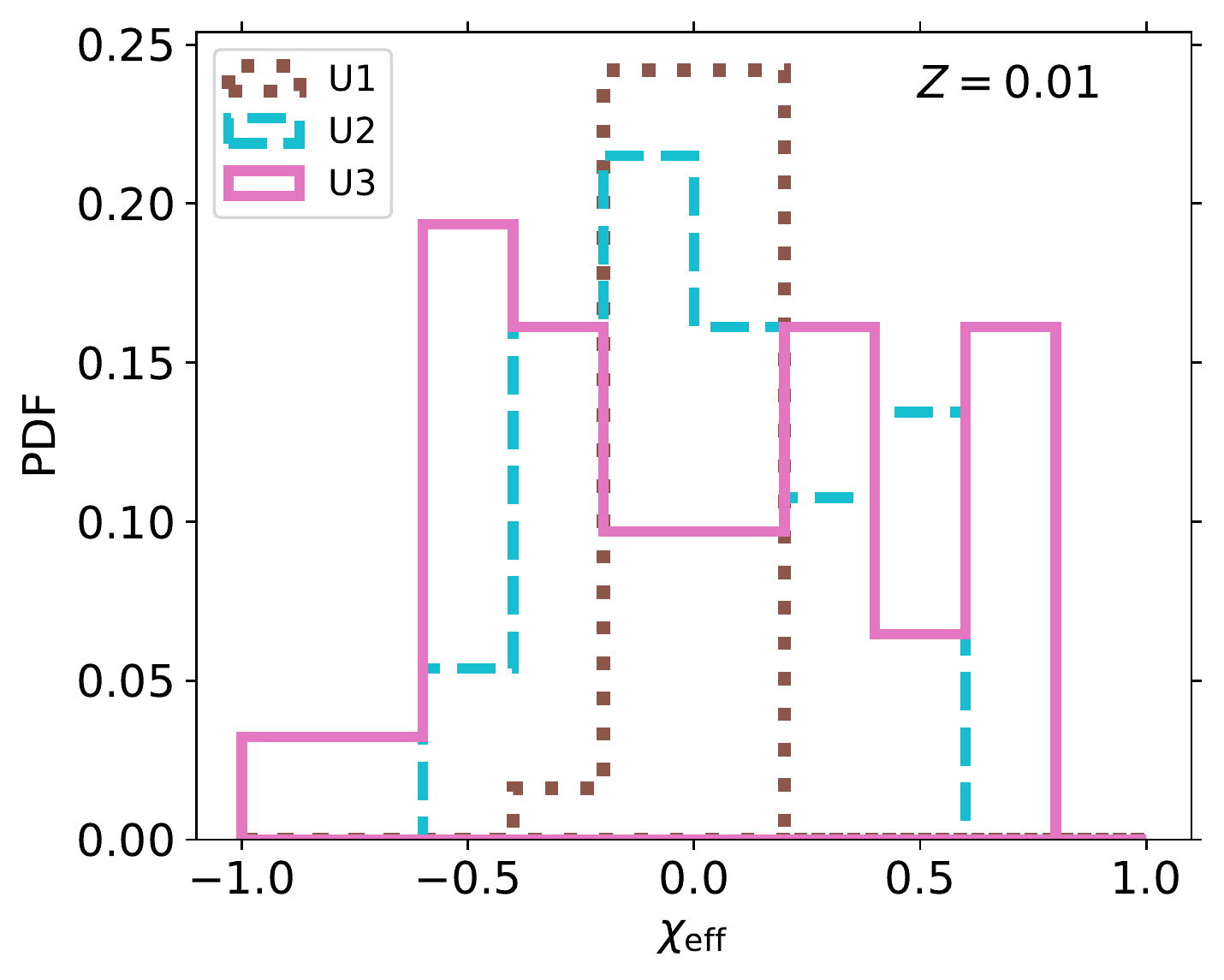}
\caption{Effective spin distributions of BH-BH binaries in triples that lead to merger for different values of $Z$ (top: Model B1; centre: Model A1; bottom: Model B4) and all the spin models under consideration (see Table~\ref{tab:spins}).}
\label{fig:spin1}
\end{figure*}

\begin{figure*} 
\centering
\includegraphics[scale=0.55]{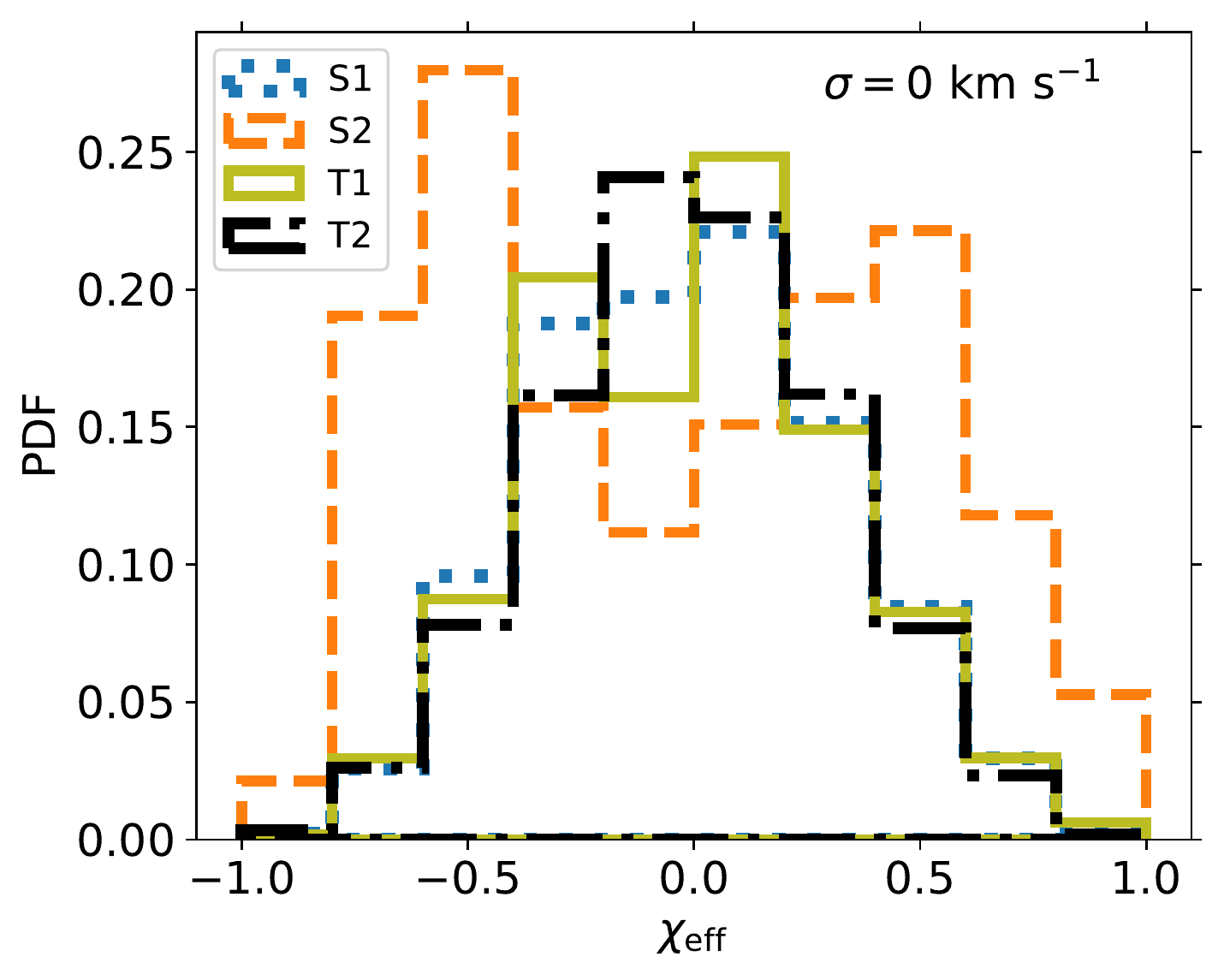}
\hspace{0.5cm}
\includegraphics[scale=0.55]{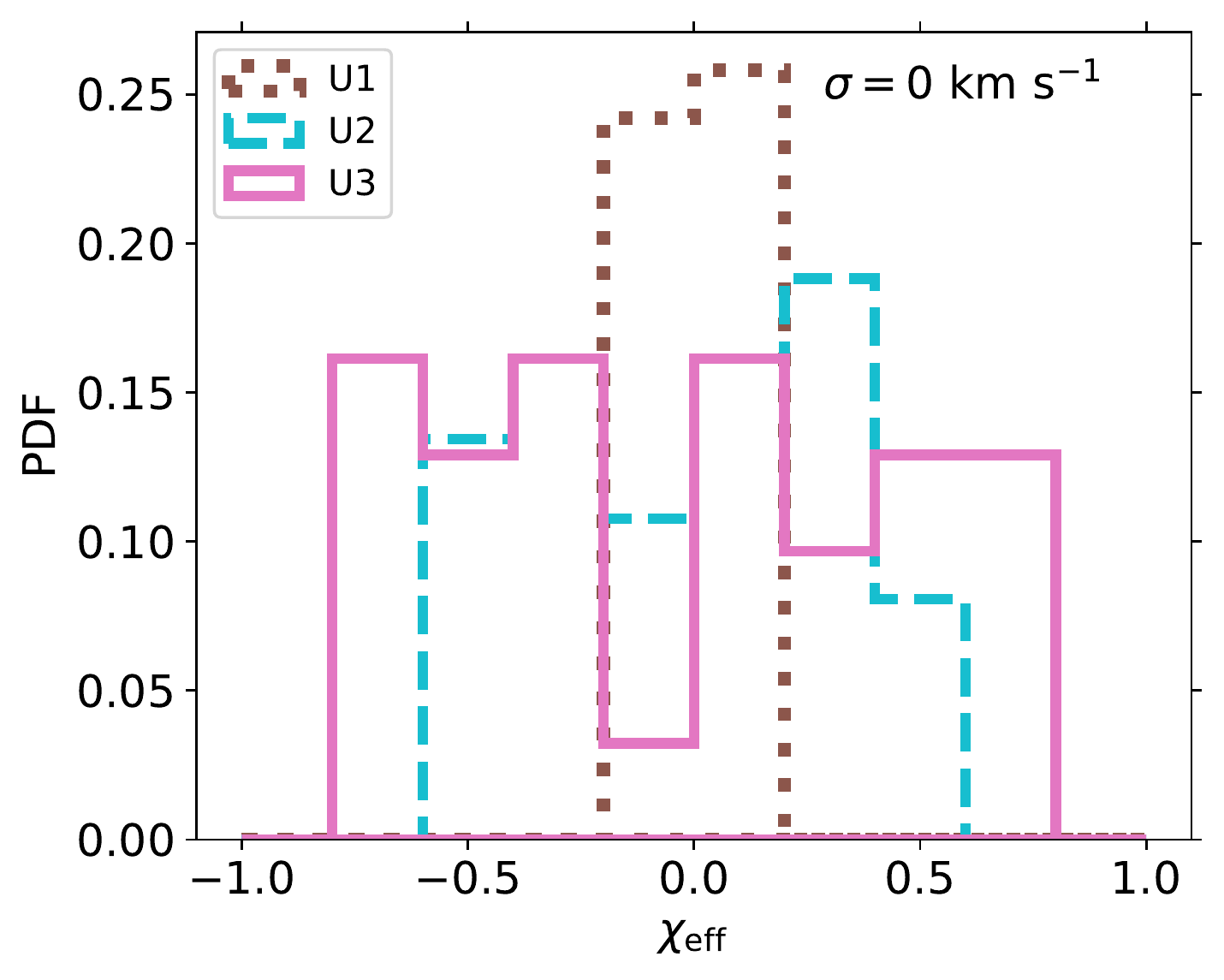}
\includegraphics[scale=0.55]{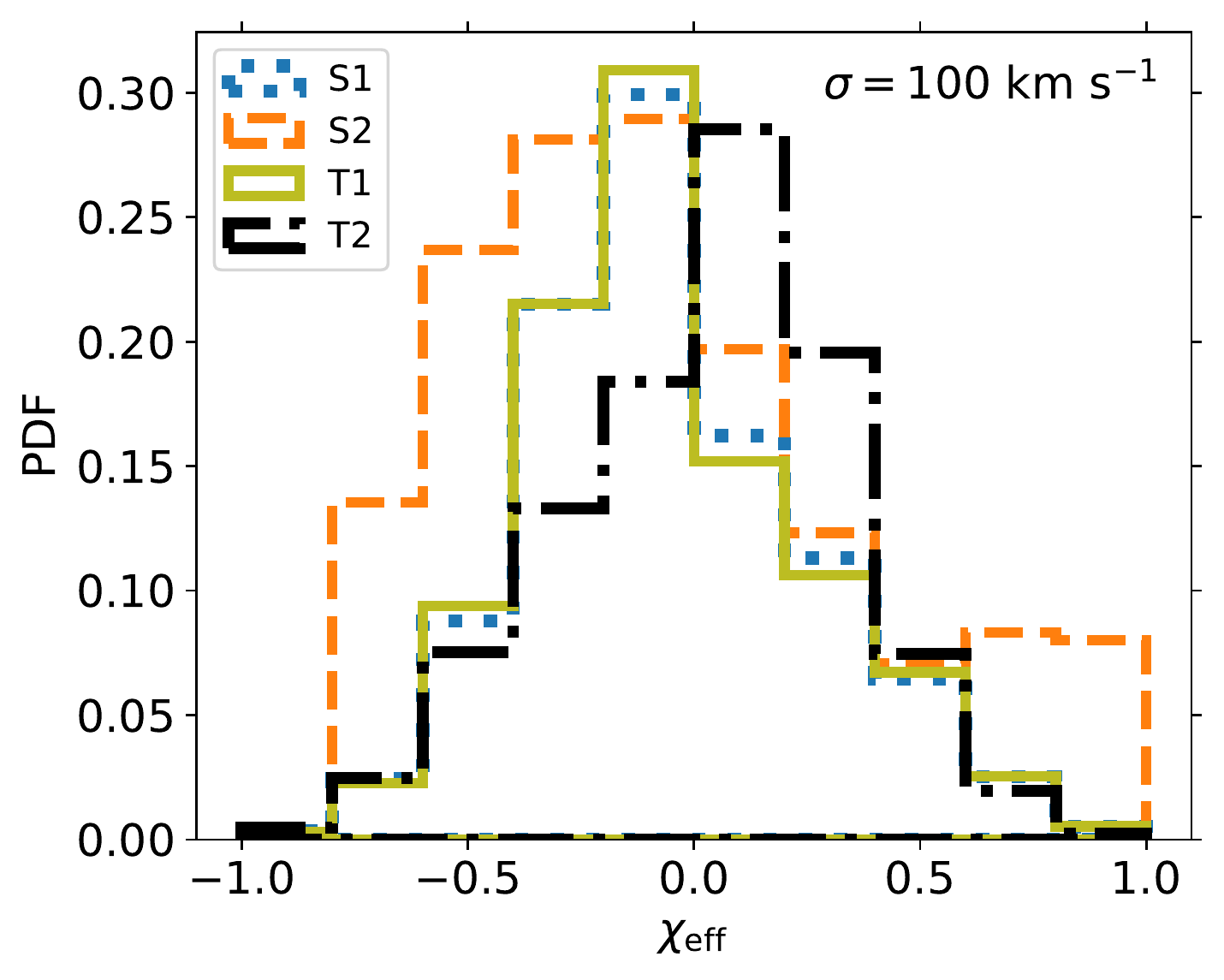}
\hspace{0.5cm}
\includegraphics[scale=0.55]{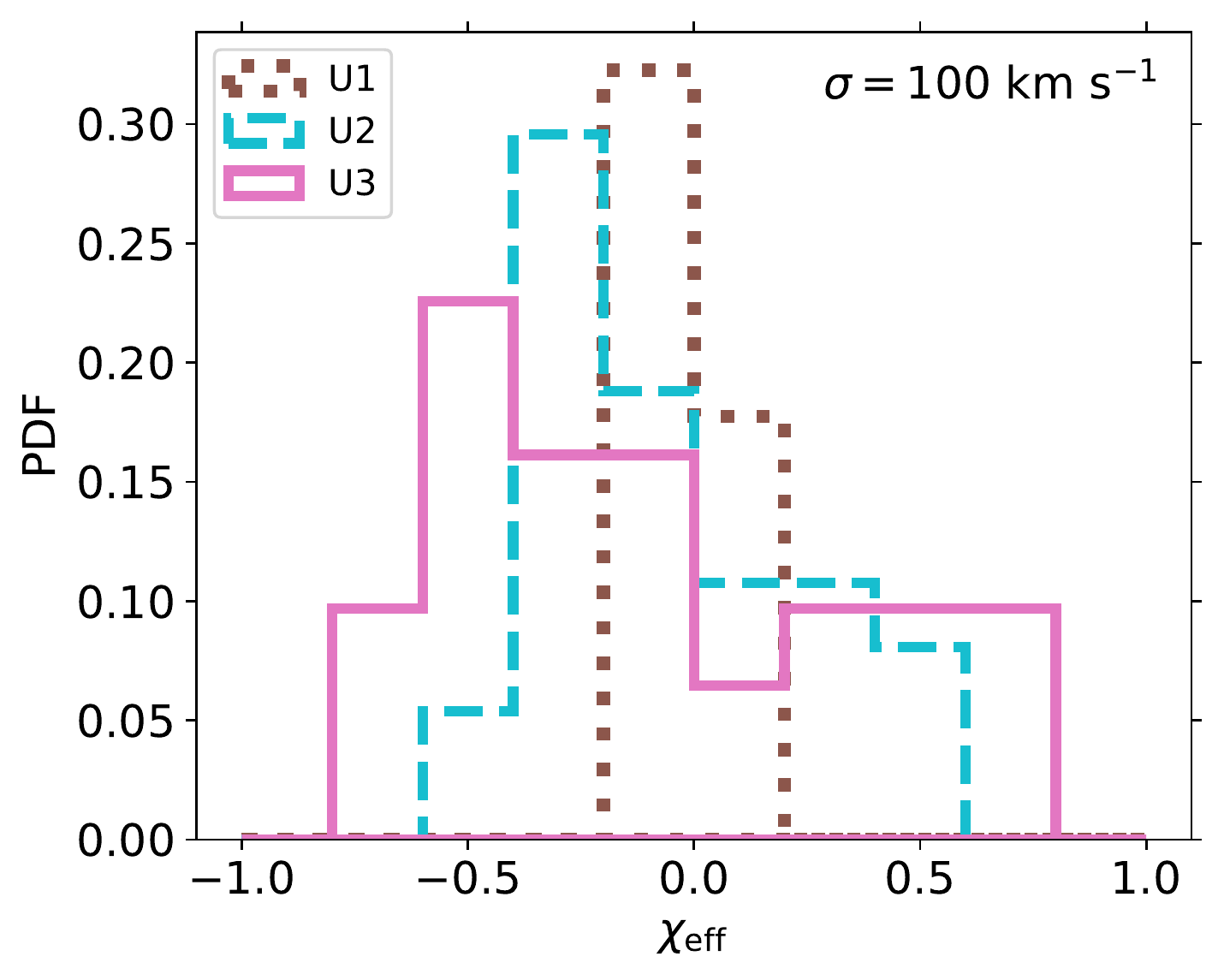}
\includegraphics[scale=0.55]{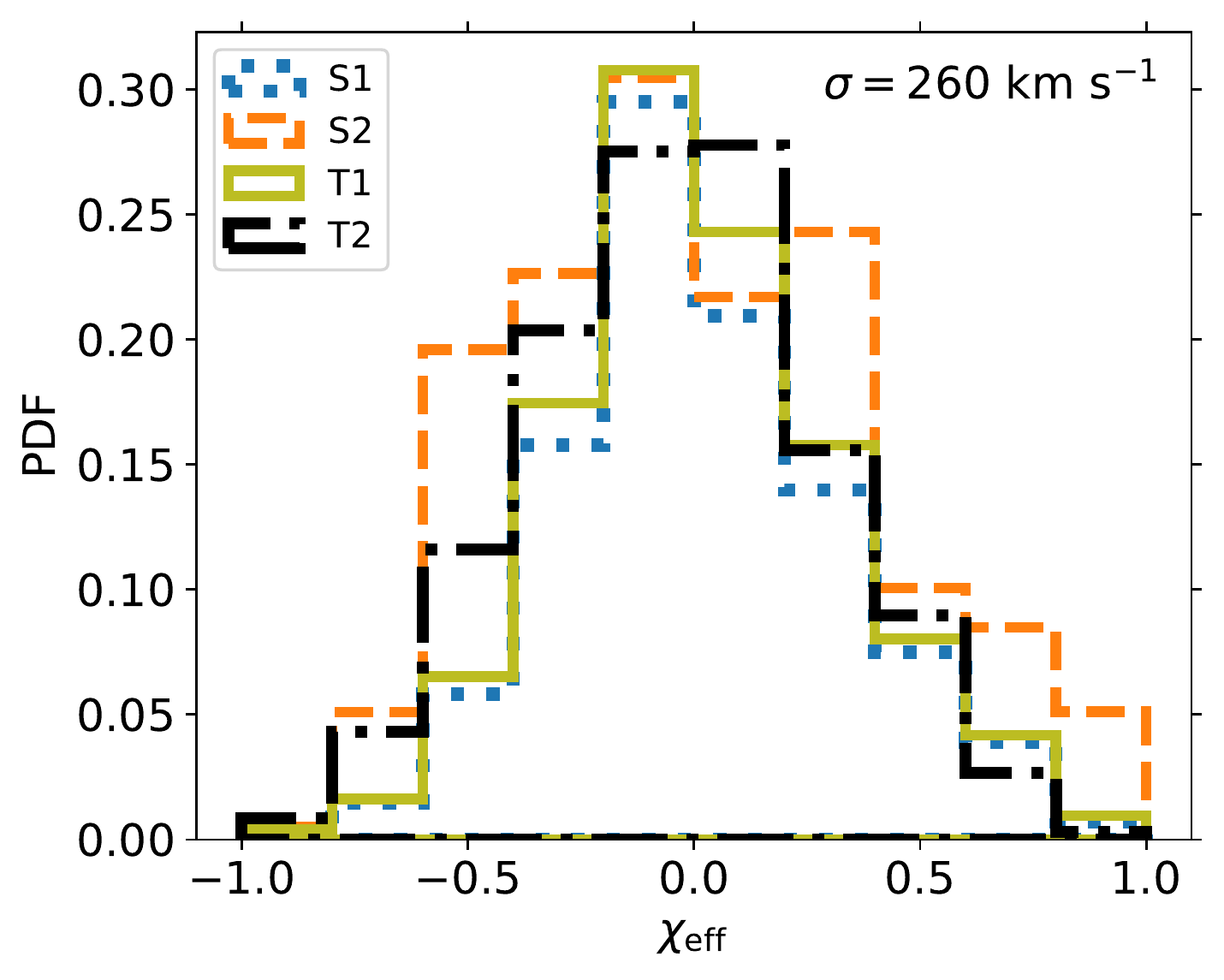}
\hspace{0.5cm}
\includegraphics[scale=0.55]{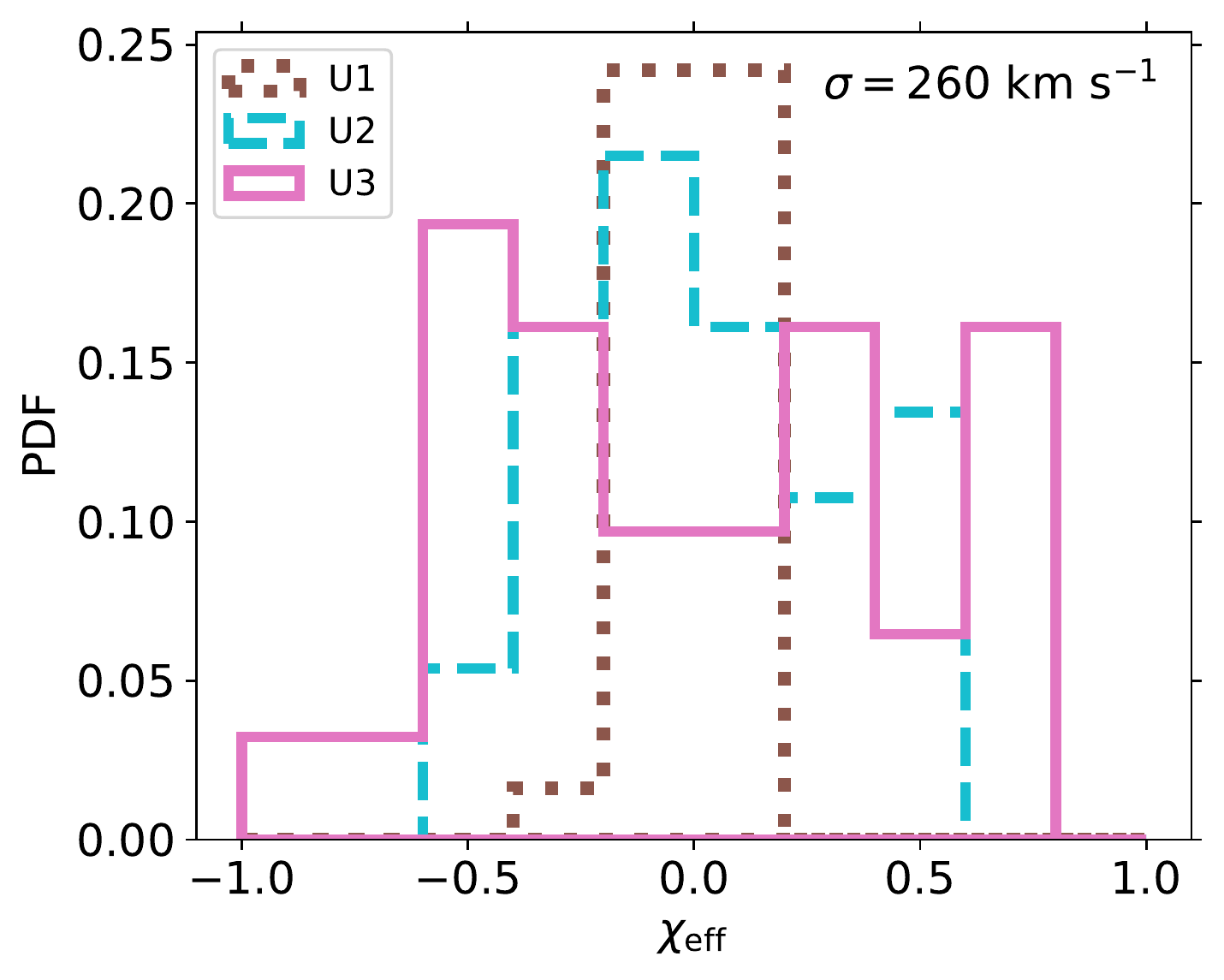}
\caption{Effective spin distributions of BH-BH binaries in triples that lead to merger for different values of $\sigma$ (top: Model A3; centre: Model A2; bottom: Model A1) and all the spin models under consideration (see Table~\ref{tab:spins}).}
\label{fig:spin2}
\end{figure*}

\begin{figure*} 
\centering
\includegraphics[scale=0.55]{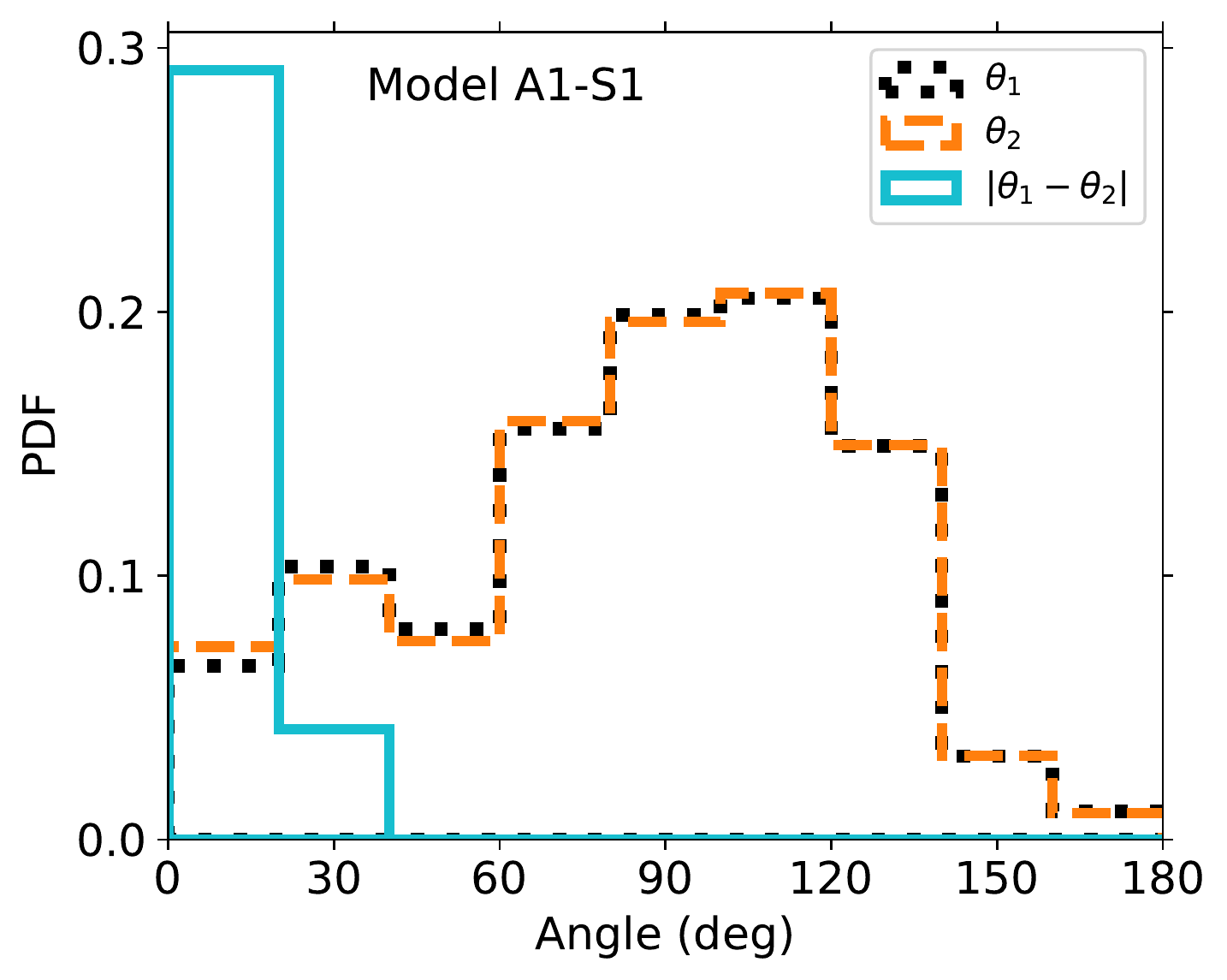}
\hspace{0.5cm}
\includegraphics[scale=0.55]{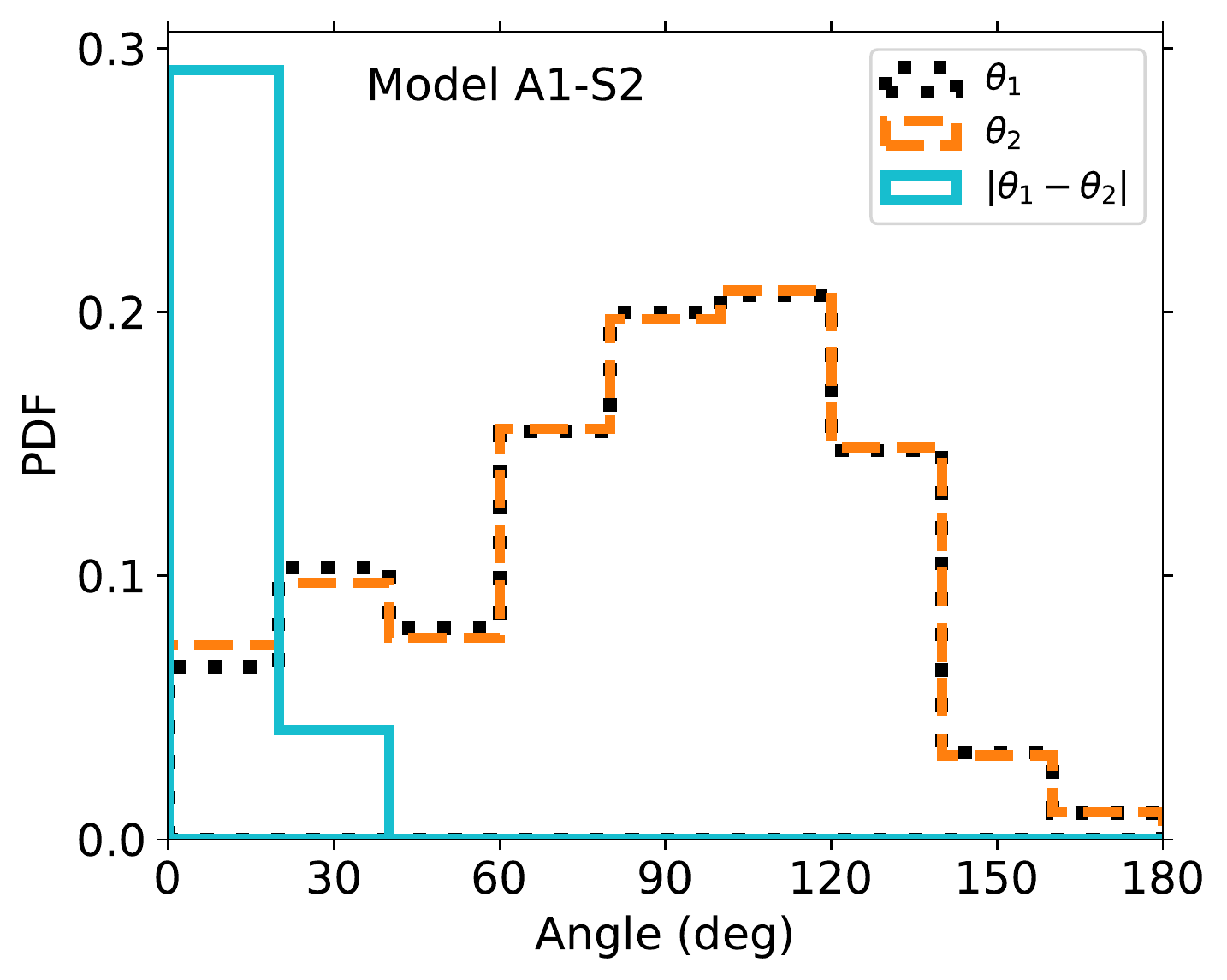}
\includegraphics[scale=0.55]{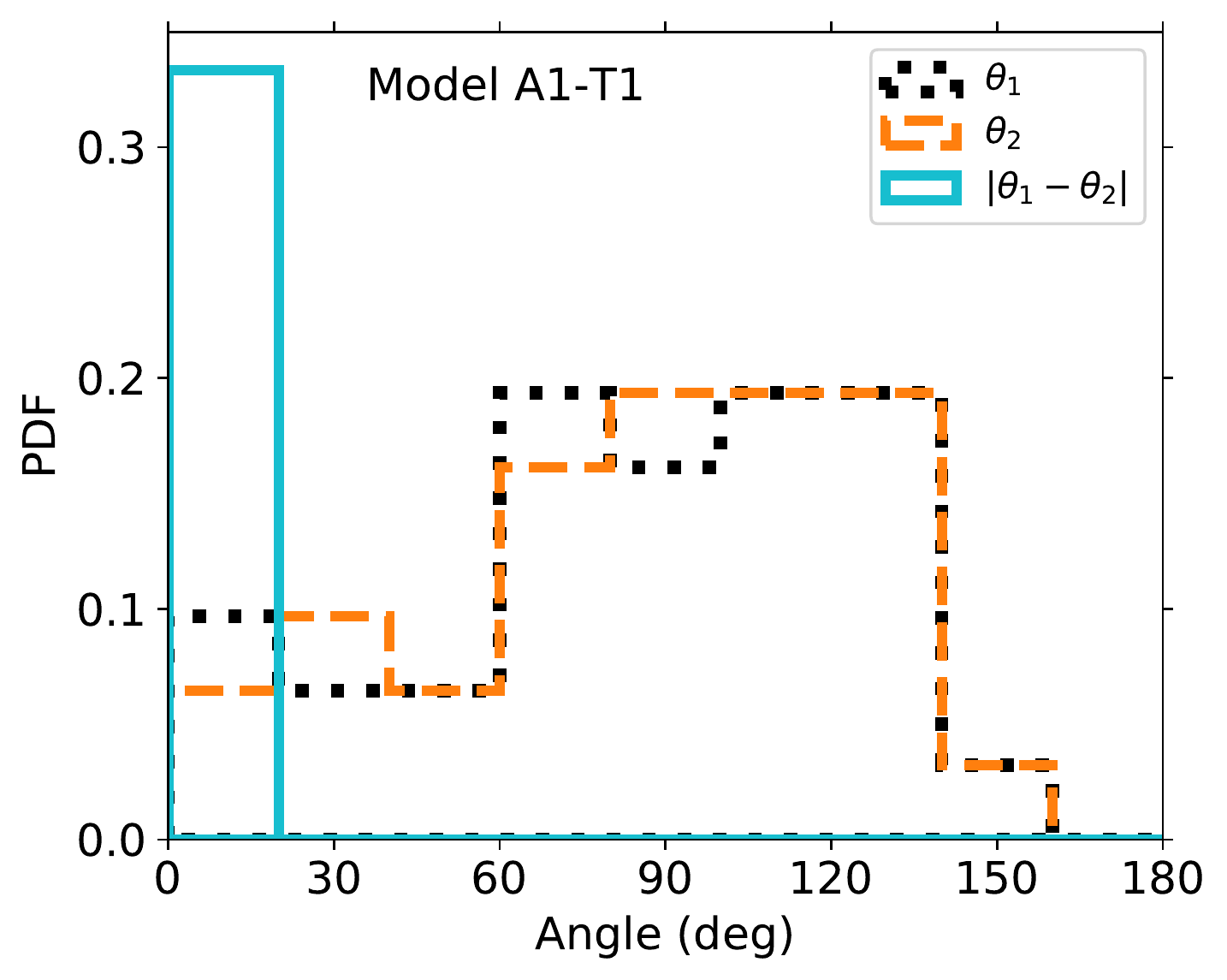}
\hspace{0.5cm}
\includegraphics[scale=0.55]{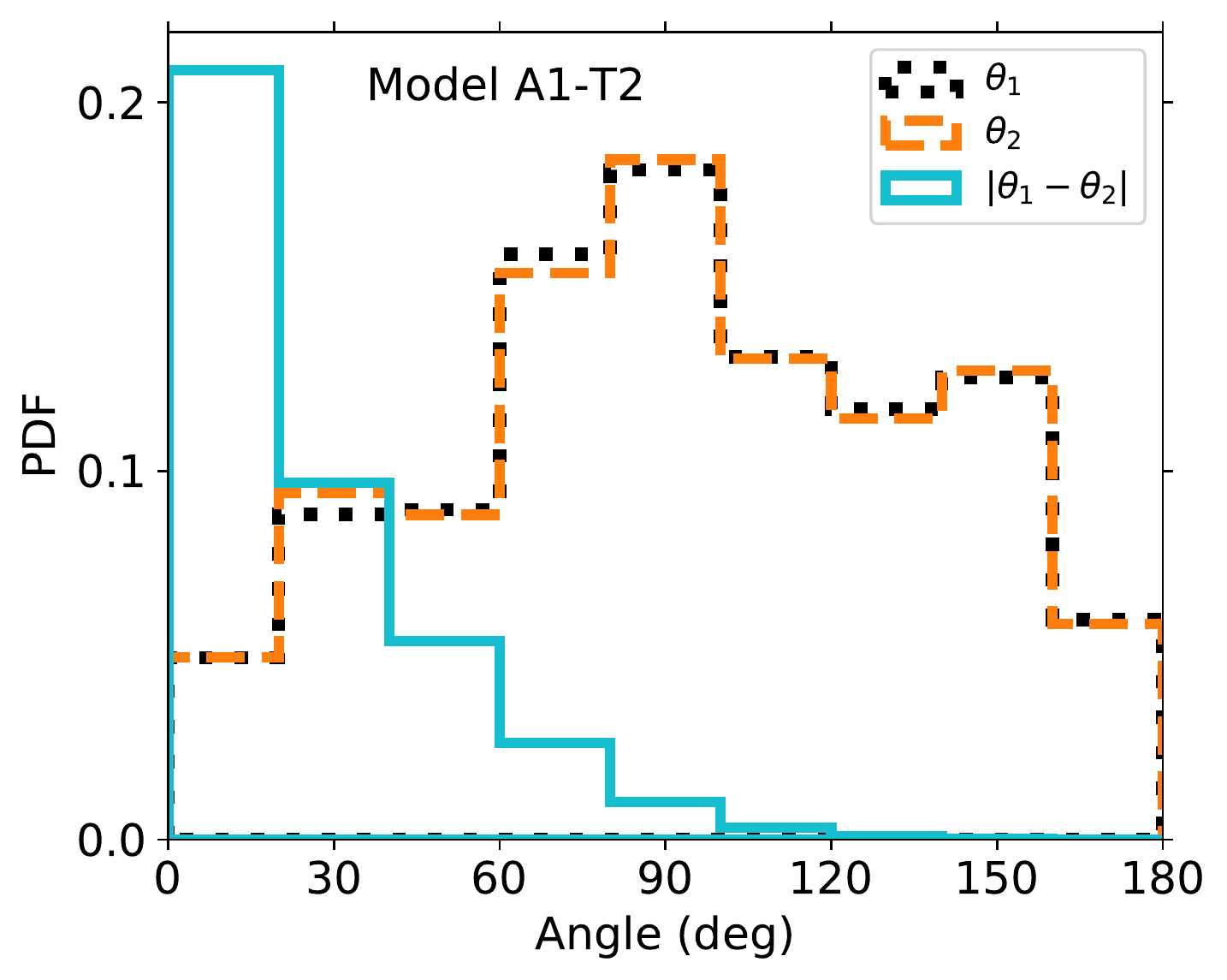}
\caption{Distributions of the absolute misalignement $\theta_1$ between ${\bf{S_{1}}}$ and $\bf{J}$, $\theta_2$ between ${\bf{S_{2}}}$ and $\bf{J}$, and the relative misalignement between ${\bf{S_{1}}}$ and ${\bf{S_{2}}}$ for Model A1 and the first four spin models of Table~\ref{tab:spins}.}
\label{fig:spin3}
\end{figure*}

The magnitudes of the BH spins are expected to be set by the physics governing the stellar collapse, which depends on the progenitor star metallicity and mass loss. However, since the specific predictions on how the progenitor properties set the remnant spin is still highly uncertain, therefore we run simulations for seven different models for the spins in order to understand how the results depend on the initial spin of the BHs. Each model differs in the initial magnitude of the Kerr spin parameter and/or the initial orientation of the spins (see Table~\ref{tab:spins})
\begin{itemize}
\item Model S1: $\chi_1$ and $\chi_2$ are drawn independently from an uniform distribution and angles between $0^\circ \le\cos\theta_{1,2}^{\rm ini}\le 20^\circ$;\footnote{Here $\cos\theta_{1,2}=\mathbf{\hat{S}}_{1,2}\cdot\mathbf{\hat{L}}$ describes the spin misalignment angle relative to the orbital angular momentum vector.}
\item Model S2: $\chi_1$ and $\chi_2$ set by the BH mass \citep{belc2017}
\begin{equation}
\chi_{1,2}=\frac{p_1-p_2}{2}\tanh\left(p_3-\frac{M_{\rm BH1,2}}{\msun}\right)+\frac{p_1+p_2}{2}\ ,
\label{eqn:bhspin}
\end{equation}
where $p_1=0.86\pm 0.06$, $p_2=0.13\pm 0.13$, and $p_3=29.5\pm 8.5$. Following \citet{gerosa2018}, spins are generated by drawing random samples uniformly in the region in between the two curves given by the upper and lower limits of the parameters. Eq.~\ref{eqn:bhspin} captures some of the key features found in \citet{belc2017}, that is heavier BHs tend to have smaller spins, and reflects the uncertainties of this model \citep[see e.g. Figure~1 in][]{gerosa2018}. The misalignment angles are drawn from $0^\circ \le\cos\theta_{1,2}^{\rm ini}\le 20^\circ$ uniformly;
\item Model T1: $\chi_1$ and $\chi_2$ are drawn independently from an uniform distribution and the spins are aligned with the inner angular momentum;
\item Model T2: $\chi_1$ and $\chi_2$ are drawn independently from an uniform distribution and the initial spin-orbit misalignents of the BHs are drawn from an isotropic distribution;
\item Models U1, U2, U3: $\chi_1$ and $\chi_2$ are fixed at $0.2$, $0.5$, $0.8$, respectively, and $0^\circ \le\cos\theta_{1,2}^{\rm ini}\le 20^\circ$ is drawn uniformly.
\end{itemize}

In Figure~\ref{fig:spin1}, we show the final $\chieff$ distributions of merging BH-BH binaries in triples for different values of the progenitor metallicity. In the right panel of Figure~\ref{fig:spin1}, we show the PDFs of the effective spins for Models U1, U2, and U3, where we fix the initial Kerr parameters of the BHs to $0.2$, $0.5$, $0.8$, respectively. We find that mergers with high BH spins have a broadly distributed $\chieff$, and mergers with low absolute spins have $\chieff$ peaked around zero. The distributions do not depend significantly on the value of $Z$. 

In the left panel, we illustrate the PDFs for Models S1, S2, T1, T2, in which the magnitude of initial spins are sampled from a broad distribution (Table~\ref{tab:spins}). For these distributions, the $\chieff$ distributions have a peak at $\chieff\sim 0$ and broad tails,
which is not significantly affected by the initial distribution of the spins. A possible exception is Model S2 for $Z=0.0001$ and $Z=0.001$. In these models, we sample $\chibh$ from Eq.~\eqref{eqn:bhspin}. Since low metallicities produce heavy BHs and Eq.~\eqref{eqn:bhspin} implies small initial spins on average, the final PDF results peak at $\chieff\sim 0$ with less important tails.

In Figure~\ref{fig:spin2}, we show the PDFs of $\chieff$ of as function of $\sigma$. The final distributions do not depend significantly on the initial choice of the spins of the BHs in Models S1, S2, T1, T2, and presents a peak at $\sim 0$ with broad tails. The only outlier is the $\chieff$ distribution in the case $\sigma=0\kms$ for Model S2, where the distribution is flatter. In the models where we fix the Kerr parameters, the effective spin PDFs do not depend on $\sigma$ and present a similar behaviour described above for different $Z$'s. In Table~\ref{tab:data}, we report the $\chieff$ inferred from observed BH-BH mergers \citep{ligo2018,venu2019,zack2019}. In general, some of the models are consistent with the observed distributions.

Finally, we show the distributions of the absolute misalignements $\theta_1$ and $\theta_2$ between ${\bf{S_1}}$ and $\bf{J}$ and ${\bf{S_2}}$ and $\bf{J}$, respectively, and $|\theta_1-\theta_2|$ in Figure~\ref{fig:spin3}. The distributions are similar to each other for the spin Models S1, S2, T1, showing that $|\theta_1-\theta_2|$ is peaked narrowly at zero, while Model T2 shows a broader tail. Note that this is due to the fact that the initial orientation of the BH spins were drawn in a correlated way for models S1 and S2 while they are drawn independently from an isotropic distribution in Model T2. The initial correlations may be due to the interaction among progenitor stars, particularly tidal dissipation. Figure~\ref{fig:spin3} shows that the relative orientation of the BH spins do not randomize during the evolution, but they retain the information on the initial conditions. However merging systems in triples do not show counteralignment with $|\theta_1-\theta_2|>90^{\circ}$ irrespective of the initial condition.

We do not find any significant correlation between the spin and the eccentricity in our different models. The only significant correlation is in model S2, where we use Eq.~\ref{eqn:bhspin}. In this case, heavier BHs tend to have smaller spins \citet{belc2017}.

\subsection{Merger rates}

\begin{figure*} 
\centering
\includegraphics[scale=0.55]{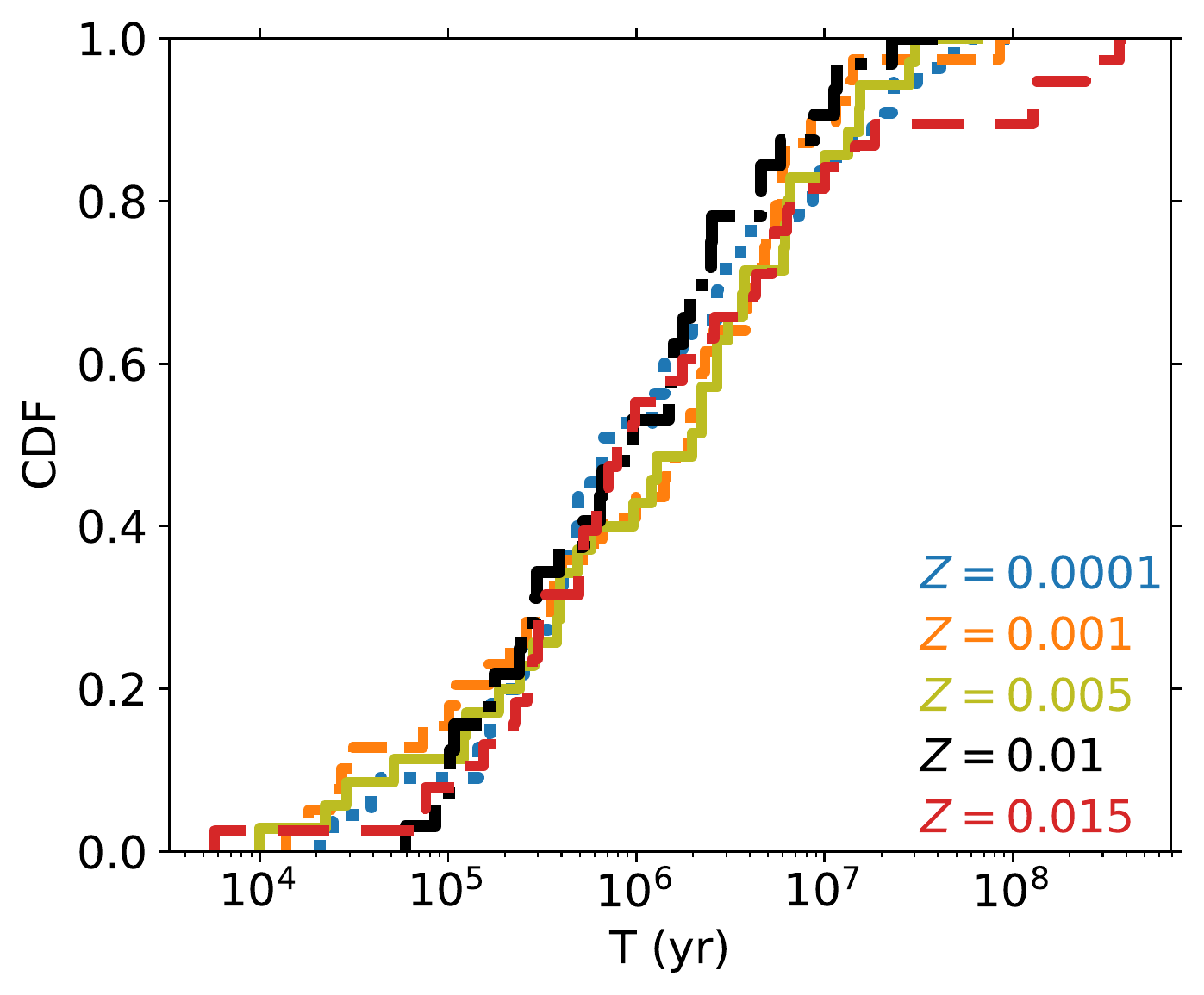}
\hspace{0.5cm}
\includegraphics[scale=0.55]{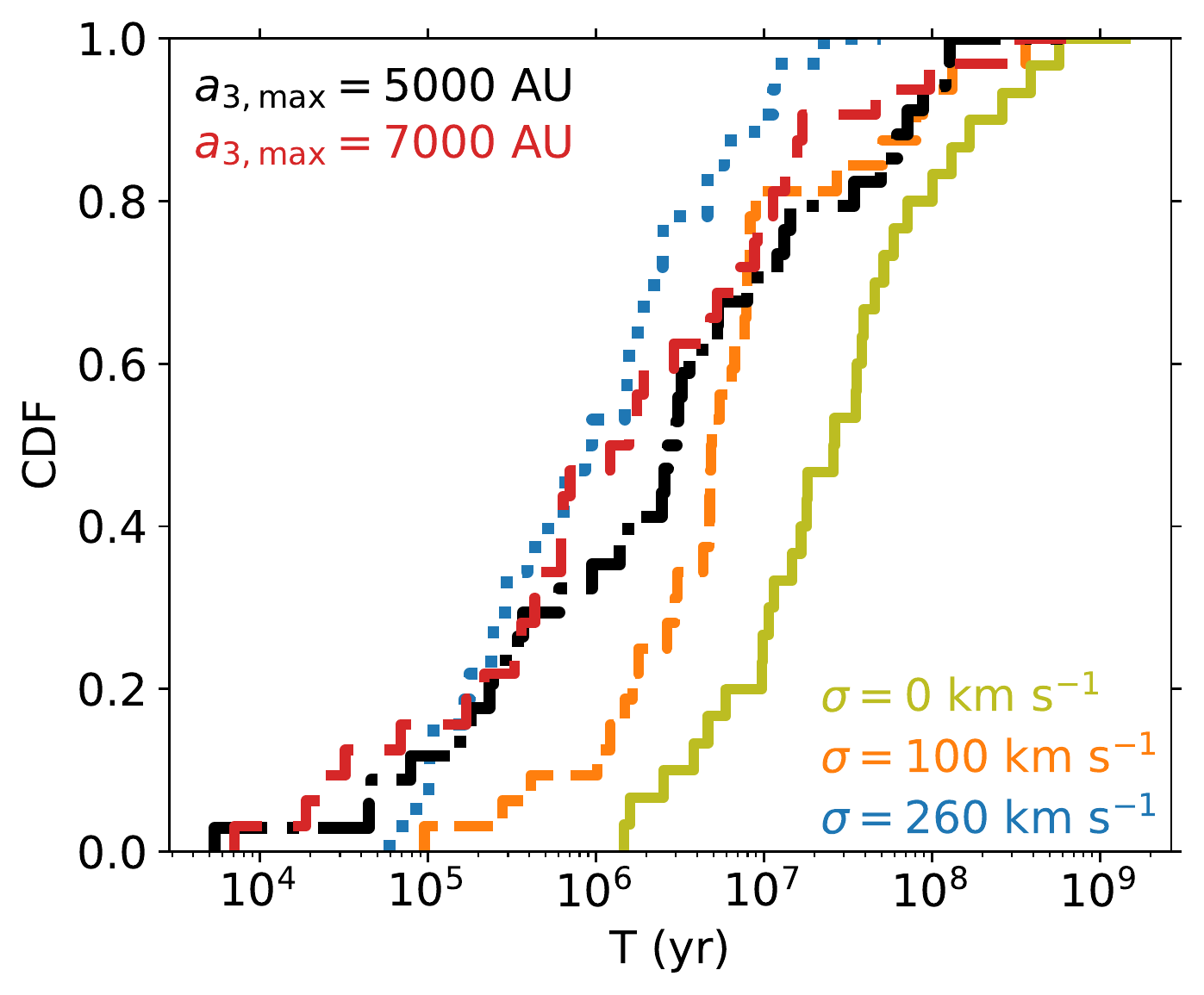}
\caption{Merger time distribution of BH-BH binaries in triples that lead to merger (see Table~\ref{tab:models}). Left panel: different progenitors metallicities and $\sigma=260\kms$; right panel: different kick velocities and triple orbital parameters and $Z=0.01$.}
\label{fig:tmerge}
\end{figure*}

Figure~\ref{fig:tmerge} shows the merger time CDFs of BH-BH binaries in triples that lead to merger for different progenitor metallicities (left; $\sigma=260\kms$) and different kick velocities and triple orbital parameters (right; $Z=0.01$). In our simulations, the distribution of merger times does not depend significantly on the progenitor metallicity nor on the possible maximum separation of the triple $\amax$. The only parameter on which the merger time depends significantly is the mean kick velocity, since larger values of $\sigma$ unbind more wider triples. Therefore, triples surviving the SNe are wider for lower $\sigma$ on average and merge on a longer timescale since their LK timescale is longer.

In order to compute the merger rate of BH-BH binaries in triples, we follow the method outlined in \citet{sil17} and \citet{flpk2019}. We assume that the local star formation rate is $0.025 \msun$ Mpc$^{-3}$ yr$^{-1}$ \citep{both2011}, thus the number of stars formed per unit mass, volume, and time is given by,
\begin{align}
\dot{n}(m) &= \frac{\eta_{\rm SFR}\, f(m)}{\langle m\rangle} =
\nonumber\\&=
5.2\times 10^6 \left(\frac{m}{\msun}\right)^{-2.3}\ \mathrm{M}_\odot^{-1}\ \mathrm{Gpc}^{-3}\mathrm{yr}^{-1}\ ,
\end{align}
where $\langle m\rangle = 0.38 \msun$ is the average stellar mass for a Kroupa mass function. Assuming a constant star-formation rate\footnote{The star-formation rate depends on the cosmic time. \citet{rodant2018} adopted a redshift-dependent star-formation rate from \citet{madau2014} and found a similar range for the overall merger rate of triple BHs.}, the BH-BH merger rate in triple systems is then,
\begin{align}
\mathcal{R}_\mathrm{BH-BH} &=
(1-\zeta) f_{\rm 3} f_{\rm stable} f_{\rm merge}\int_{20\msun}^{150\msun} \dot{n}(m_1) dm_1=
\nonumber\\&=
7.4\times 10^4 \eta (1-\zeta) f_{\rm 3} f_{\rm stable} f_{\rm BH-BH}\ \mathrm{Gpc}^{-3}\ \mathrm{yr}^{-1}\ ,
\label{eqn:rrr}
\end{align}
where $f_{\rm 3}=0.25$ is the fraction of stars in triples, $f_{\rm stable}$ is the fraction of systems that remain stable after the SN events take place, $f_{\rm merge}$ is the fraction of systems that merge (see Table~\ref{tab:models}), and $\eta$ is the conditional probability that the secondary is also a progenitor of a BH given that the primary is a BH progenitor:
\begin{equation}
    \eta = \frac{ \int_{20\msun}^{150\msun} d m_1 f_{\rm IMF}(m_1) \int_{{{20\msun}/{m_1}}}^{1} dq_{12}  f_{q}(q_{12}) 
    }{
     \int_{20 \msun}^{150\msun} d m_1 f_{\rm IMF}(m_1) }\,,
\end{equation}
where $f_q(q_{12})$ is the mass ratio distribution of the inner binary, which we assume to be constant. For a \citet{kroupa2001} initial mass function $f_{\rm IMF}$ we get $\eta= 0.38$. In Eq.~\eqref{eqn:rrr}, $f_{\rm merge}\sim 0.06$ for all the models, we find that the fraction of stable systems depends both on the mean natal kick and on $Z$, because lower progenitor metallicities lead to more massive BHs, which are on average imparted lower kicks at birth as a result of our assumption of momentum-conserving kicks (see Tab.\ref{tab:models}). A factor of uncertainty is the possible KL dynamics during the evolution of the stellar triples before they form a BH-BH system in the inner binary, which we have not modeled here. Some fraction of the parameter space can be removed by the earlier evolution of the system \citep{shapp2013}. To estimate this uncertainty, we assume very conservatively that any stellar triple whose initial KL timescale is less than the lifetime of the primary star \citep[$\sim 7$ Myr;][]{iben91,hurley00,maeder09} in the inner binary will merge, and, as a consequence, will not form a triple system with an inner BH-BH binary \citep{rodant2018}. We find that the fraction of these triples is $\zeta\sim 0.83$ on average, except for Model A2 where we find $\zeta\sim 0.55$. Using the minimum and maximum values of $f_{\rm stable}$ in Table~\ref{tab:models} and plugging numbers into Eq.~\eqref{eqn:rrr},
\begin{equation}
\Gamma_\mathrm{BH-BH}=0.008-9 \ \mathrm{Gpc}^{-3}\ \mathrm{yr}^{-1}\ .
\end{equation}
This is consistent with the merger rates in triples derived by \citet{ant17} and \citet{sil17}.
The higher merger rates correspond to low natal kicks and low progenitor metallicities. For a log-uniform distribution of mass ratios, we estimate a rate $\sim 2$ times larger.

\section{Discussion and conclusions}
\label{sect:conc}

Many astrophysical scenarios have been proposed to explain the BH and NS mergers observed via GW emission by the LIGO-Virgo collaboration. A promising way to disentangle the contributions from different channels is to statistically compare the distributions of masses, spins, eccentricity and redshift of the merging binaries as discussed in Section~\ref{sect:intro}.
In this paper, we have studied the dynamical evolution of BH triples in isolation by means of high-precision $N$-body simulations, including post-Newtonian terms up to order 2.5PN. We started from the main sequence progenitors of the BHs and modeled the SNe that lead to the formation of the BH triple. We adopted different prescriptions for the SN natal kicks, and considered different progenitor metallicities and orbital parameters. We have shown that the typical eccentricity of BH-BH binaries merging in triple systems when reaching the 10 Hz is $0.01$-$0.1$ and that the merger rate is in the range $0.02-24 \ \mathrm{Gpc}^{-3}\ \mathrm{yr}^{-1}$, depending on the natal kick prescriptions and progenitor metallicity, consistent with the rates inferred by \citet{ant17} and \citet{sil17}. Higher rates correspond to low natal kicks and low metallicities. We find that the fraction of mergers with a detectable nonzero eccentricity for the LIGO--VIRGO-KAGRA network at design sensitivity is in the range $\sim 9$--$42\%$.

We confirm the findings of \citet{liulai2019} that the BH spin exhibits a wide range of evolutionary paths, and different distributions of final spin-orbit misalignments can be produced depending on the system parameters. The effective spin parameter $\chieff$ is broadly distributed if the merging BHs are highly spinning. Nevertheless, $\chieff$ is peaked at zero for all triple population models that we have investigated, due to the fact that many merging systems have low intrinsic BH spins in these models, consistent with \citet{antonini2018} and \citet{rodant2018}. We have also discussed that the triple scenario we studied in this paper could reproduce the distribution of effective spins inferred from LIGO-Virgo mergers and the IAS group \citep{ligo2018,venu2019,zack2019}.

We note that when we check that the triple systems remain stable after each SN event, systems that are deemed unstable by the \citet{mar01} criterion could still merge, thus possibly enhancing our inferred merger rates. In our calculations, we assumed that the SN events take place instantaneously and do not simulate the systems during the main sequence lifetime of the progenitors. Nevertheless, we have used fits to single stellar evolutionary tracks to determine the final BH mass \citep{spera2015}. However, the details of the specific evolutionary paths, which depend on stellar winds, metallicity and rotation, of the stellar progenitors could reduce the parameter space \citep{shapp2013}. The situation becomes more complicated if mass loss during possible episodes of Roche-lobe overflows and common evolution phases in the triple are considered \citep{rosa2019,hamd2019}. We do not model these possible effects and leave the detailed study to future investigations.

Ongoing and future observations promise to observe hundreds of merging BHs, thus providing a sufficiently high number to constrain the different formation scenarios. The GW observation of a merging BH-BH binary which enters the LIGO band with a non-negligible eccentricity and with a nearly zero effective spin would be consistent with triple scenario. 

While our results show that the mass-weighted sum of the spin vectors projected on the orbital angular momentum vector, $\chieff$, is approximately symmetrically distributed around $\sim 0$ in the triple channel, we find that the angle between the individual spins is strictly less than $\sim 90^\circ$ (Figure~\ref{fig:spin3}). While this parameter is poorly constrained in the O1 and O2 observing runs, it is potentially measurable through the effects of spin-precession in future longer duration detections at lower GW frequencies \citep[see also][]{Khan2019,Fairhurst2019}. The misalignment angle between the two spins represents a powerful constraint to test the triple channel of BH mergers.

\section*{Acknowledgements}

GF thanks Seppo Mikkola for helpful discussions on the use of the code \textsc{archain}, and Barak Zackay for providing data on the effective spins of binary black hole mergers. We thank Fabio Antonini for useful discussions. GF acknowledges support from a CIERA postdoctoral fellowship at Northwestern University. This work received funding from the European Research Council (ERC) under the European Union’s Horizon 2020 Programme for Research and Innovation ERC-2014-STG under grant agreement No. 638435 (GalNUC) and from the Hungarian National Research, Development, and Innovation Office under grant NKFIH KH-125675 (to BK). This research was supported in part by the National Science Foundation under Grant No. NSF PHY-1748958. 

\bibliographystyle{mn2e}
\bibliography{refs}

\end{document}